\documentclass[usegraphicx,usenatbib]{mn2e}
\usepackage{times}

\renewcommand\[{\begin{equation}}
\renewcommand\]{\end{equation}}

\catcode`\@=11
\def\gsim{\ifmmode{\mathrel{\mathpalette\@versim>}}
    \else{$\mathrel{\mathpalette\@versim>}$}\fi}
\def\lsim{\ifmmode{\mathrel{\mathpalette\@versim<}}
    \else{$\mathrel{\mathpalette\@versim<}$}\fi}
\def\@versim#1#2{\lower 2.9truept \vbox{\baselineskip 0pt \lineskip
    0.5truept \ialign{$\m@th#1\hfil##\hfil$\crcr#2\crcr\sim\crcr}}}
\catcode`\@=12
\def\Micm{M_{\rm gas}}
\def\Mzicm{\Micm^Z}
\def\xv{{\bf x}}
\def\Zicm{Z}
\def\ZicmM{<\hspace{-0.05in}\Zicm\hspace{-0.05in}>}
\def\TicmM{<\hspace{-0.05in}T\hspace{-0.05in}>}
\def\ZicmL{<\hspace{-0.05in}\Zicm\hspace{-0.05in}>_L}
\def\TicmL{<\hspace{-0.05in}T\hspace{-0.05in}>_L}
\def\Zpr{Z_{\rm pr}}
\def\Tpr{T_{\rm pr}}
\def\SigX{\Sigma_{\rm X}}
\def\Lx{L_{\rm X}}
\def\Emis{{\cal{E}}}
\def\Tvir{T_{\rm vir}}
\def\rc{r_{\rm c}}
\def\rg{r_{\rm g}}
\def\rvir{r_{\rm vir}}
\def\rt{r_{\rm t}}
\def\phiz{\phi_0}
\def\rhoz{\rho_0}
\def\Tz{T_0}
\def\betaz{\beta_0}
\def\phit{\phi_{\rm t}}
\def\rhot{\rho_{\rm t}}

\def\Tt{T_{\rm t}}
\def\Tzt{T_{\rm 0t}}
\def\Ttr{T_{\rm cr}}
\def\pz{p_0}
\def\rz{r_{\rm Z}}
\def\Tspec{T_{\rm sl}}
\def\mH{m_{\rm H}}
\def\Msun{M_{\odot}}

\def\Psit{\Psi_{\rm t}}

\def\Zsun{Z_{\odot}}

\def\mz{m_{\rm Z}}

   \title[Average $Z$ and $T$ for hydrostatic gas distributions]
         {Hydrostatic gas distributions: global estimates
of temperature and abundance }

   \author[Ciotti and Pellegrini]
          {Luca Ciotti and Silvia Pellegrini
           \\ Astronomy Department, University of Bologna, 
              via Ranzani 1, 40127 Bologna, Italy
           }

\date{Accepted 2008 April 08. Received 2008 March 28; in original form 2008 January 08}
\pubyear{2008}

\begin{document} 
\maketitle

\begin{abstract} 

  Estimating the temperature and metal abundance of the intracluster
  and the intragroup media is crucial to determine their global metal
  content and to determine fundamental cosmological parameters. When a
  spatially resolved temperature or abundance profile cannot be
  recovered from observations (e.g., for distant objects), or
  deprojection is difficult (e.g., due to a significant non-spherical
  shape), only global average temperature and abundance are derived.
  After introducing a general technique to build hydrostatic gaseous
  distributions of prescribed density profile in potential wells of
  any shape, we compute the global mass weighted and emission weighted
  temperature and abundance for a large set of barotropic equilibria
  and an observationally motivated abundance gradient. We also compute
  the spectroscopic-like temperature that is recovered from a single
  temperature fit of observed spectra.  The derived emission weighted
  abundance and temperatures are higher by 50\% to 100\% than the
  corresponding mass weighted quantities, with overestimates that
  increase with the gas mean temperature. Spectroscopic temperatures
  are intermediate between mass and luminosity weighted temperatures.
  Dark matter flattening does not lead to significant differences in
  the values of the average temperatures or abundances with respect to
  the corresponding spherical case (except for extreme cases).

\end{abstract}


\section{Introduction}
\label{secint}

The amount of metals in the Intracluster Medium (ICM) and in the
Intragroup Medium (IGM) gives us important clues about the past star
formation activity of the stellar population of these galaxy systems,
being it directly linked to the total number of supernovae exploded in
the past and to the initial stellar mass function of the star
formation epoch (e.g., Renzini et al. 1993).  The metal content 
can also enlight how the enrichment proceeded, e.g., via
stripping or galactic winds driven by SNe or AGN feedback, and has 
implications for both the ICM/IGM and galaxy evolution (e.g., Wu,
Fabian, \& Nulsen 2000; Finoguenov et al. 2001; Kapferer et al. 2007).
For these reasons the observational study of the metal content of the
ICM/IGM is growing fastly.  After the first large compilation of
(emission weighted) average abundance values of iron from $EXOSAT$,
$Einstein$ and $GINGA$ observations (Arnaud et al. 1992), $ASCA$
made metal measurements for many clusters (Fukazawa et al. 1994,
Finoguenov et al. 2000, Baumgartner et
al. 2005).  The average iron abundance was estimated to be $0.38\pm
0.07$ and $0.21\pm 0.05$ respectively for the cooling flow and non
cooling flow clusters (Allen \& Fabian 1998). In more recent times,
the superior quality of the $XMM-Newton$ and $Chandra$ instrumentation
has allowed for more accurate determinations of the elemental
abundance pattern (e.g., Tamura et al. 2004, Fukazawa et al. 2004,
Durret et al. 2005, Sanders \& Fabian 2006, de Plaa et al. 2007,
Finoguenov et al. 2007, Rasmussen \& Ponman 2007).  Nowadays, these
studies are carried on also with $Suzaku$ (e.g., Matsushita et
al. 2007, Sato et al. 2007).

Similarly to the metal abundance, the hot ICM/IGM temperature is also
one of the most important and commonly used global observables: it is
used as a proxy for the total mass of the system (e.g., Voit 2005),
from which the clusters can be used as probes for fundamental
cosmological parameters (e.g., Henry \& Arnaud 1991, Henry 1997,
Nevalainen et al. 2000, Arnaud et al. 2005). Temperature profiles have
been built with improved quality in the recent past (e.g., Arnaud et al. 2005,
Pointecouteau et al. 2005, Vikhlinin et al. 2005, 2006, Pratt et al.
2007, Rasmussen \& Ponman 2007).  Since the ICM/IGM are not
isothermal, ideally the mass weighted temperature should enter the
computation of quantities to be used for cosmological tests.

From a more quantitative point of view, 
the amount of the mass of metals in the ICM/IGM is given by  
\[
\Mzicm=\int\rho (\xv)\Zicm (\xv)d^3\xv,
\label{eqmzicm}
\]
where $\rho$ and $\Zicm$ are the true three dimensional gas density
and abundance profiles. Thus, the mass weighted average abundance is
given by
\[
\ZicmM ={\Mzicm\over\Micm},
\label{zicmm}
\]
where $\Micm =\int\rho (\xv)d^3\xv$ is the total hot gas mass.
Similarly, the mass weighted average temperature is
\[
\TicmM ={\int\rho(\xv) T(\xv) d^3\xv\over \Micm}.
\label{tmass}
\]
Unfortunately, there are at least three serious problems with
estimating $\ZicmM$ and $\TicmM$ from observations: 1) for many
clusters/groups we do not know the {\it intrinsic shape} of the gas
distribution and the {\it viewing angles} under which we are observing
it; therefore, one cannot uniquely deproject observed quantities
(obtained in general from X-ray data) to derive $\rho$, $T$, and
$\Zicm$; 2) even for spherically symmetric systems, deprojection is a
demanding numerical process, very sensitive to the properties of the
instrumental PSF and to measurement errors (e.g., Finoguenov \& Ponman
1999); 3) in many cases only a single spectrum can be extracted for
the whole gas, and only an average abundance and temperature can be
obtained; this happens when there are not enough counts for a
spatially resolved spectroscopy, e.g., for distant clusters/groups
(Hashimoto et al. 2004, Maughan et al. 2007,
Baldi et al. 2007 for recent observations with {\it Chandra} and
XMM-$Newton$).  In particular, the average abundance and temperature
mentioned in point 3) above are {\it not} those given in
eqs.~(\ref{zicmm})-(\ref{tmass}), but are in practice luminosity
weighted quantities (e.g., Mathiesen \& Evrard 2001, Mazzotta et
al. 2004, Maughan et al. 2007, Rasia et al. 2005, Kapferer et
al. 2007) that can be defined as
\[
\ZicmL={\displaystyle{\int\SigX(\xi_1,\xi_2)\Zpr(\xi_1,\xi_2)d\xi_1 d\xi_2 
                       \over \Lx} },
\label{eqzl}
\]
and
\[
\TicmL={\displaystyle{\int\SigX(\xi_1,\xi_2) \Tpr(\xi_1,\xi_2)d\xi_1 d\xi_2 
                       \over \Lx} },
\label{eqtl}
\]
where $(\xi_1,\xi_2)$ are the coordinates of the projection plane,
$\SigX$ is the X-ray ICM surface brigthness, $\Zpr$ and $\Tpr$ are the
luminosity weighted projected abundance and temperature, and $\Lx
=\int\SigX d\xi_1 d\xi_2$ is the total X-ray luminosity
(see Appendix A1).

It is then natural to investigate the relation between the quantities
in eqs.~(2)-(3) and (4)-(5).  For example, Rasia et al. (2008), using
mock $XMM-Newton$ spectra for a sample of simulated clusters, find
that the iron abundance inferred from such spectra is very close to
the projection of the emission weighted values of $Z$ (i.e.,
$\ZicmL$), at least for thermal components of $kT>3$ keV and $kT<2$
keV. Kapferer et al. (2007), again using simulations, similarly find
that for $kT>3$ keV the X-ray emission weighted abundance is close
within few percents to that derived from the analysis of synthetic
X-ray spectra.  Unfortunately, neglecting a possible spatial variation
of the metal abundance can lead to largely wrong estimates of $\Mzicm$
when using $\ZicmL$ instead of $\ZicmM$ in eq.~(\ref{zicmm}) (Arnaud
et al. 1992).  In fact, iron distributions peaked towards the
cluster/group center have been revealed in many cases (Fukazawa et
al. 2000, Ettori et al. 2002, Sanders \& Fabian 2002, Matsushita et
al. 2003, B\"ohringer et al. 2004, Tamura et al. 2004).  Motivated by
this, in an exploratory study Pellegrini \& Ciotti (2002) showed that
in these cases $\ZicmM$ can be significantly smaller than $\ZicmL$.
Successively, De Grandi et al. (2004) confirmed this result for their
sample of cooling core clusters, for which they estimated $\ZicmM$ to
be $\sim 15$\% smaller than $\ZicmL$.

It is also well accepted that the ICM/IGM have a temperature structure
that was established by gravitational and non-gravitational processes,
as radiative cooling and heating by active galactic nuclei (see
Borgani et al. 2005, Vikhlinin et al. 2005, Piffaretti et al. 2005,
Arnaud et al. 2005, Donahue et al. 2006).  Efforts have been made
recently to understand the meaning of the temperature derived from
spectroscopic observations when the ICM/IGM has a complex thermal
structure (Mazzotta et al. 2004, Rasia et al. 2005, Vikhlinin 2006,
Nagai et al. 2007). Mazzotta et al. (2004) found that the observed
temperature, recovered from a single temperature fit to the spectrum
of a plasma with components at different temperatures (but all
continuum-dominated, i.e., with $kT\gsim 3$ keV) and extracted from
$Chandra$ or $XMM-$Newton data, is well approximated by a
"spectroscopic-like temperature" $\Tspec$ (see Sect.~\ref{res}).
Vikhlinin (2006) extended this previous work and proposed an
algorithm to accurately predict $\Tspec$ that would be derived for a
plasma with components in a wider range of temperatures ($kT\gsim 0.5$
keV) and arbitrary abundances of heavy elements.  From the analysis
of mock spectra of simulated clusters, it was found that $\Tspec$ is
lower than the emission weighted temperature $\TicmL$, with
consequences for using the observed $M-T$ relation to infer the
amplitude of the power spectrum of primordial fluctuations (Rasia et
al. 2005).

Here, extending the preliminary discussion of Pellegrini \& Ciotti
(2002) based on spherical models, we estimate how much discrepant
$\ZicmL$ and $\ZicmM$, and $\TicmL$ (or $\Tspec$) and $\TicmM$ are, by
using different plausible profiles for $\rho$, $T$ and $Z$ obtained
assuming hydrostatic equilibrium within triaxial mass distributions
resembling real systems.  In particular the models are constructed by
using a technique that allows for building analytical barotropic
gas distributions with prescribed density profiles departing from
spherical symmetry. These new models extend the class of equilibria
usually considered in the literature beyond isothermal or polytropic
models (i.e., Suto, Sasaki \& Makino 1998; Pellegrini \& Ciotti 2002;
Lee \& Suto 2003, 2004; Ostriker, Bode \& Babul 2005; Ascasibar \&
Diego 2007). In the computation of the averages, our approach takes
also advantage of the Projection Theorem, from which it follows that
$\ZicmL$ and $\TicmL$ are independent of the specific direction of the
line-of-sight, and can be calculated using the intrinsic
three-dimensional quantities of the models, with a much easier
procedure that avoids projection and surface integration.

The paper is organized as follows. In Section 2 we present the
models of the dark matter halos and the procedure to build fully
analytical hydrostatic configurations in potentials of triaxial shape,
for gas distributions corresponding to truncated quasi-isothermal
models, quasi-polytropic models and modified $\beta$ models. In
Section 3 we describe the results and in Section 4 we summarize the
main conclusions; technical results are reported in the Appendix.

\section{The models}
\label{secmet}

\subsection{Density profiles for the gravitating mass}
\label{clumod}

The density of the (dark) mass distribution is the generalization to the 
triaxial case of the so-called
$\gamma$-models (Dehnen 1993, Tremaine et al. 1994): 
\[
\varrho = {M(3-\gamma)\over 4\pi \rc^3(1-\epsilon)(1-\eta)}
{1\over m^{\gamma}(1+m)^{4-\gamma}},
\label{rhocl}
\]
where
\[
m^2 = {x^2\over\rc^2}+{y^2\over (1-\epsilon)^2\rc^2}
                     +{z^2\over (1-\eta)^2\rc^2},
\]

$M$ is the total dark mass of the system, $\rc$ is a characteristic
scale, and the pair $(\epsilon, \eta)$ parameterizes the flattening
along the $y$ and $z$ axes respectively. The mass distribution is
spherically symmetric when $\epsilon=\eta=0$, and $M$ remains constant
for different choices of the flattening. For simplicity, we restrict
to the $\gamma=0$ and the $\gamma=1$ cases: in the former, the density
profile shows a central ``core'', while in the latter the Hernquist
(1990) profile is recovered in the spherical limit.  Note that the
$\gamma=1$ models have the same radial trend, in the central regions,
as the profile obtained from high resolution cosmological simulations
(Dubinsky \& Carlberg 1991; Navarro, Frenk \& White 1996), while they
are steeper at large radii ($\propto r^{-4}$ instead of $\propto
r^{-3}$). Even though not required by the technique described in
Sect. 2.2, in our analysis we used the potential profiles obtained by
means of homeoidal expansions of the true potential at fixed total
mass (e.g., Muccione \& Ciotti 2003, 2004; Lee \& Suto 2003, 2004;
Ciotti \& Bertin 2005, hereafter CB05).  This approach has the
advantage of avoiding the numerical integration needed to recover the
potential (e.g. Binney \& Tremaine 2008), and the formulae obtained
are a very good approximation of the exact potential associated with
eq.~(\ref{rhocl}).

\begin{figure*}
\includegraphics[height=.36\textheight,width=.48\textwidth]{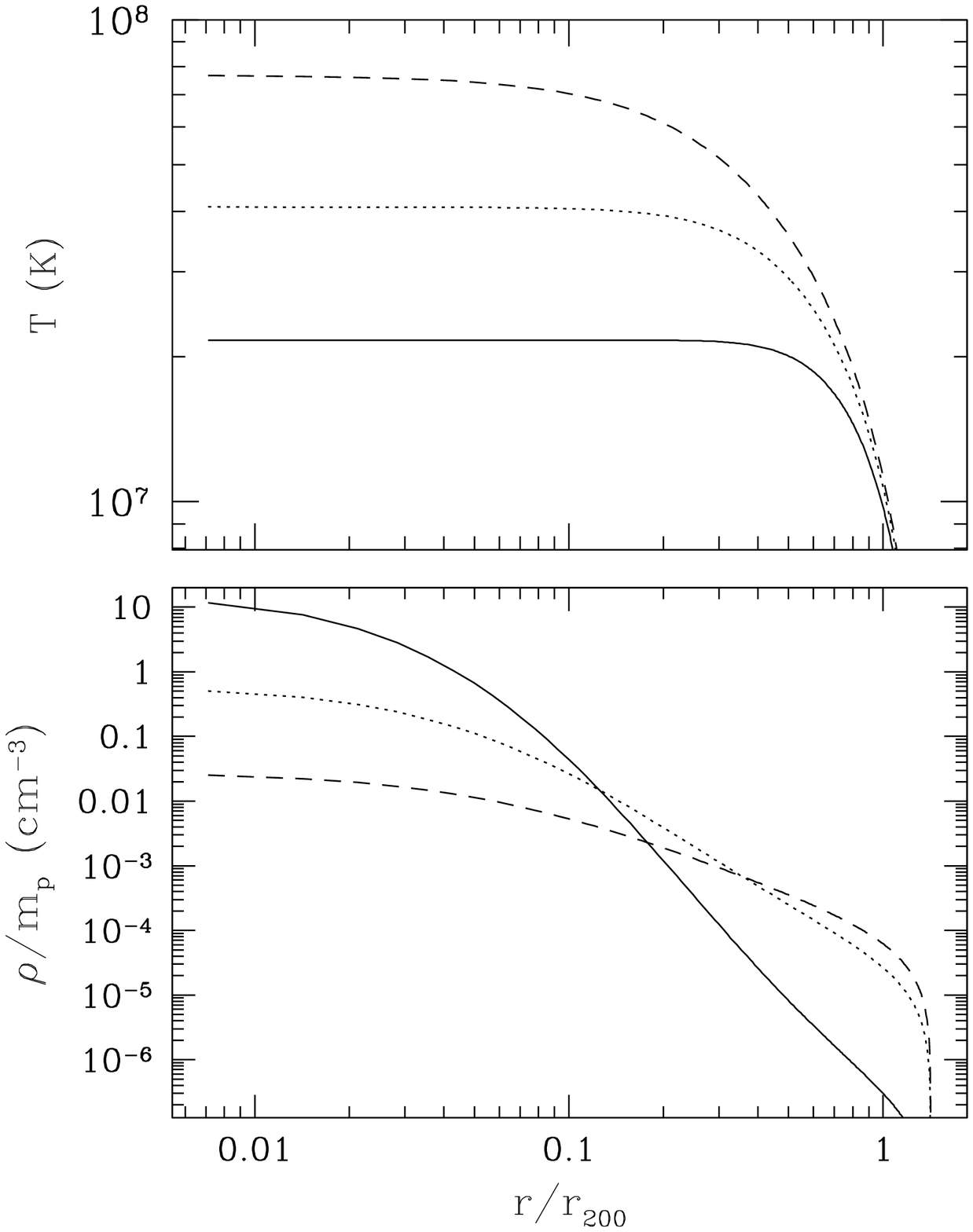}
\includegraphics[height=.36\textheight,width=.48\textwidth]{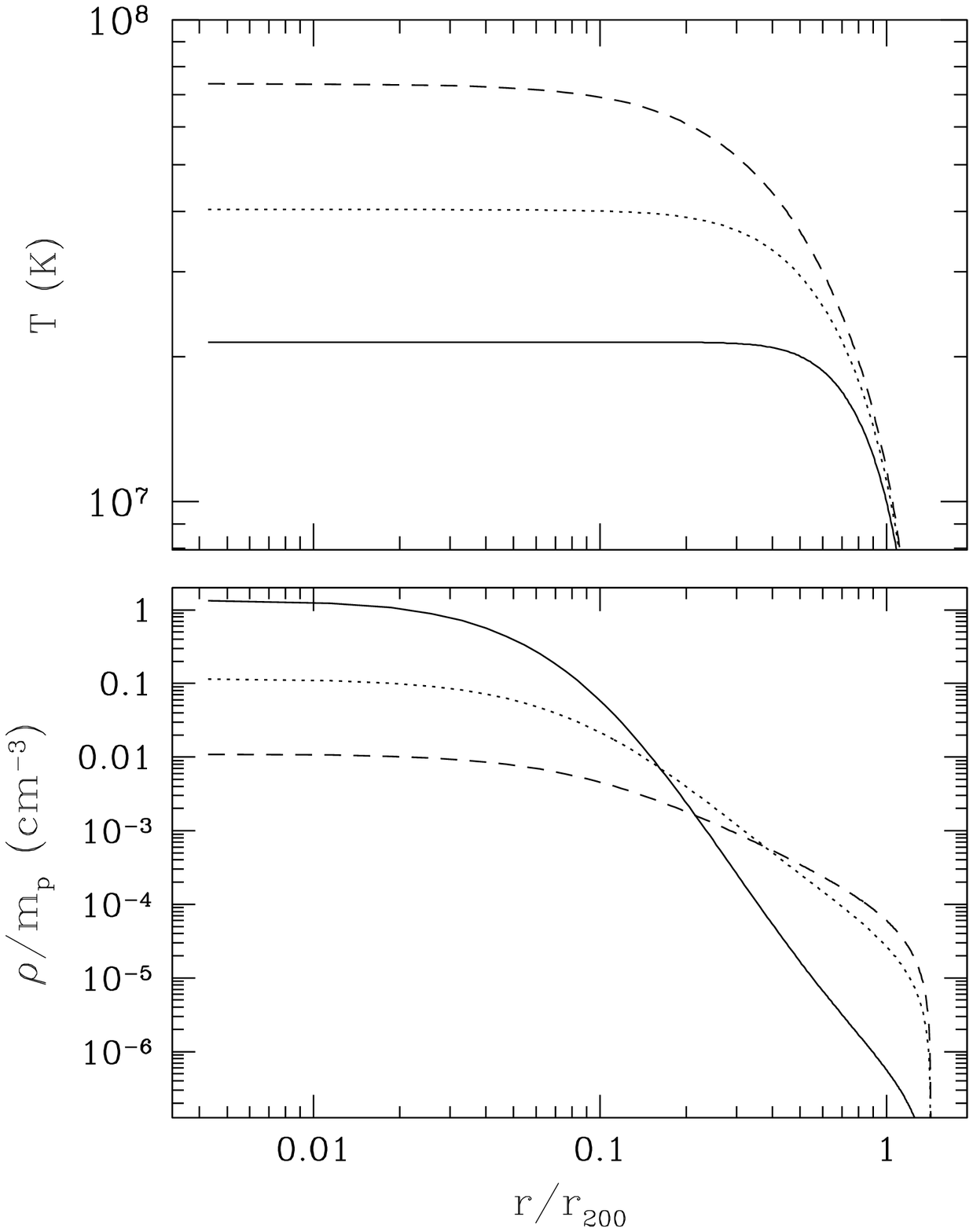}
\caption{Radial profiles of density and temperature for
truncated quasi-isothermal spherical gas models 
in the Hernquist potential (left, $\rt=\rvir =6\rc$) and in the $\gamma=0$ 
potential (right, $\rt=\rvir=10\rc$), with $\Tvir=2.3\,{\rm keV}$. 
Three values of $\Tz$ of the parent isothermal model have been 
chosen: $\Tz=0.8\,\Tvir$ (solid lines), $\Tz=1.5\,\Tvir$ (dotted lines)
and  $\Tz=3\,\Tvir$ (dashed lines).
The distance from the center $r$ is normalized to $r_{200}$ as
defined at the end of Sect.~\ref{clumod}.}
\label{iso}  
\end{figure*}

Homeoidal expansion applied to the $\gamma =0$ model shows that
\begin{eqnarray}
\tilde\phi =-{2r+1\over 2(r+1)^2}-
            \left[{3r^2+12r+8\over 2(r+1)^2}-{4\ln (r+1)\over r}\right]
            {\epsilon+\eta\over r^2}-\nonumber\\
            \left[{12\ln (r+1)\over r}-{3r^3+22r^2+30r+12\over (r+1)^3}
            \right]{\epsilon y^2+\eta z^2\over r^4},\quad\quad
\label{phicz}
\end{eqnarray}
where $\phi=GM\tilde\phi/\rc$, and the value of the central potential
is $\tilde\phiz =-(3+\epsilon+\eta)/6$. For the $\gamma=1$ model
\begin{eqnarray}
\tilde\phi =  
            -{1\over r+1} -
            \left[{r+2\over r+1}-{2\ln (r+1)\over r}\right]
                  {\epsilon+\eta\over r^2}-\nonumber\\
            \left[{6\ln(r+1)\over r}-{2r^2+9r+6\over (r+1)^2}
            \right]{\epsilon y^2+\eta z^2\over r^4},
\label{phico}
\end{eqnarray}
and $\tilde\phiz =-(3+\epsilon+\eta)/3$. In the formulae above the
radial coordinates are normalized to $\rc$, and in both cases the
expansion holds for $1\geq 3\eta-\epsilon$ (see Appendix A in CB05).  
Thus, in principle the maximum deviation from spherical
symmetry is obtained for $\eta=\epsilon=0.5$, corresponding to a
prolate system of axis ratio 2:1\footnote{As shown in CB05, for 
large flattenings the
expanded density deviates from an ellipsoid, being more similar to a
toroid; however the shape of the equipotential surfaces is very
similar to that of ellipsoidal systems.}. 
Finally, the virial temperature of the system (defined as $3
k M\Tvir\equiv |U|$, where $U$ is the gravitational energy) in the
limit of small flattenings, and independently of the specific density
profile $\varrho (m)$, is given by
\[
\Tvir = {GM \over 3 \rvir}{\mu\mH \over k}
          \left(1+{\epsilon+\eta \over 3} \right),
\label{tempv}
\]
where $\mu$ is the mean particle weight, $\mH$ is the proton mass, $k$
is the Boltzmann constant and $\rvir$ is the virial radius of
$\varrho$ in the spherical limit (Muccione \& Ciotti 2004). Here 
$\rvir=10\rc$ and $6\rc$ for the $\gamma=0$ and $\gamma =1$
models, respectively.  Note that, for fixed $M$ and $\rvir$, $\Tvir$
increases for an increasing flattening. 

Summarizing, the potential is determined by assigning the two
flattenings $\epsilon$ and $\eta$, and by choosing the mass $M$, the
slope $\gamma$, and $\rc$.  The latter step is done via the relation
$\rvir= \rvir (M)$ holding for dark matter halos obtained from
cosmological simulations in a flat $\Lambda$CDM cosmological model
($\Omega_m=0.3$, $\Omega_{\Lambda} =0.7$, $h=0.7$, where the Hubble
constant is defined as $100 h$ km s$^{-1}$ Mpc$^{-1}$), as derived,
e.g., by Lanzoni et al. (2004).  For example, for a mass $M=3.5\times
10^{14} h^{-1}\Msun$ we adopt $\rvir=1.4 h^{-1}$ Mpc, so that
$\rc=0.14 h^{-1}$ Mpc for the $\gamma=0$ model, and $\rc=0.23 h^{-1}$
Mpc for the Hernquist model, with $\Tvir=2.3$ keV (spherical
case). For $M=1.0\times 10^{15} h^{-1}\Msun$, $\rvir=1.8 h^{-1}$ Mpc
and $\Tvir = 5.1$ keV (spherical case).  We also derived the commonly
used $r_{200}$ and $r_{500}$ radii (within which the average mass
density is respectively 200 and 500 times the critical density at
redshift zero for a flat $\Lambda$CDM cosmological
model). Independently of $\gamma=0$ or $\gamma=1$, $r_{200}\simeq
0.7\rvir$ and $r_{500}\simeq 0.5\rvir$ for $M=3.5\times 10^{14}
h^{-1}\Msun$, and $r_{200}\simeq 0.8\rvir$ and $r_{500}\simeq
0.6\rvir$ for $M=1.0\times 10^{15} h^{-1}\Msun$.  Remarkably, the
ratios $r_{200}/\rvir$ and $r_{500}/\rvir$ are very similar to those
typical of the Navarro et al. (2006) profile of same total mass and
virial radius.

\subsection{The hydrostatic equilibrium models}

Once a dark matter distribution is chosen, we build hydrostatic
equilibrium models for the gas within it, assuming that the gas mass
does not contribute to the gravitational field, and that the gas is
perfect so that its pressure is $p=k\rho T/\mu\mH$. Our procedure is
based on the well known result that pressure, density and temperature
in hydrostatic equilibrium are all stratified over isopotential
surfaces (e.g., Tassoul 1980)\footnote{If $\mu$ varies, it is actually
the ratio $T/\mu$ to be stratified over the isopotential surfaces, but
here we neglect the very small $\mu $ variations due to the adopted
abundance gradients.}.  In other words, hydrostatic configurations are
{\it barotropic}, i.e. $p=p(\rho)$, which allows us to solve the
hydrostatic equation $\nabla p=-\rho\nabla\phi$ for potentials of {\it
general} shape. Therefore, the method is fully general: the only
additional simplifying assumption is that the potential has a finite
minimum $\phiz$ at the center and vanishes at infinity.  With this
method we could also study the effect of substructures by
superimposing different, off-centered dark-matter halos.

\subsubsection{Truncated quasi-isothermal models}
\label{truiso}

The following is a family of exact equilibria that generalizes the
classical isothermal models
\[ 
\rho =\rhoz\;{\rm e}^{-{\phi-\phiz\over\betaz}},\quad
                  \betaz\equiv {k\Tz\over\mu\mH},
\label{rhois}
\]
where $\rho$ is the isothermal equilibrium stratification 
of temperature $\Tz$ in a
generic potential $\phi$, and $\phiz$ and $\rhoz$ are (for example)
the central potential and the central gas density. As usual for
isothermal equilibria the total mass diverges, and a truncation
surface (outside which $\rho=0$) must be introduced.  This should be
done preserving the barotropicity of the distribution. In practice,
the truncation surface must be an isopotential
surface\footnote{Note the analogy with stationary truncated stellar
systems where, according to the Jeans theorem, the truncation surface must 
be defined in terms of the
isolating integrals of the motion. At the
truncation surface, the normal component of the velocity dispersion
tensor (the temperature analogous) 
vanishes (e.g. Ciotti 2000).}.  In addition, to avoid unphysical
density jumps, it is natural to truncate the system 
by subtracting to eq.~(\ref{rhois}) (the {\it parent} distribution), 
its value on some isopotential surface $\phit$, so we consider the new density
distribution 
\[
\rho =\rhoz\; {\rm e}^{{\phiz\over\betaz}}
              \left({\rm e}^{-{\phi\over\betaz}}-
              {\rm e}^{-{\phit\over\betaz}}\right),\quad \phi\leq\phit ,
\label{rhoist}
\]
while the quasi-isothermal equilibrium temperature associated with
eq.~(\ref{rhoist}) is obtained from eq.~(A6) as
\[
{T\over\Tz}=1-{\phit -\phi\over \betaz
              \left({\rm e}^{{\phit-\phi\over\betaz}}-1\right)}.
\label{tempist}
\]

\begin{figure*}
\includegraphics[height=.36\textheight,width=.48\textwidth]{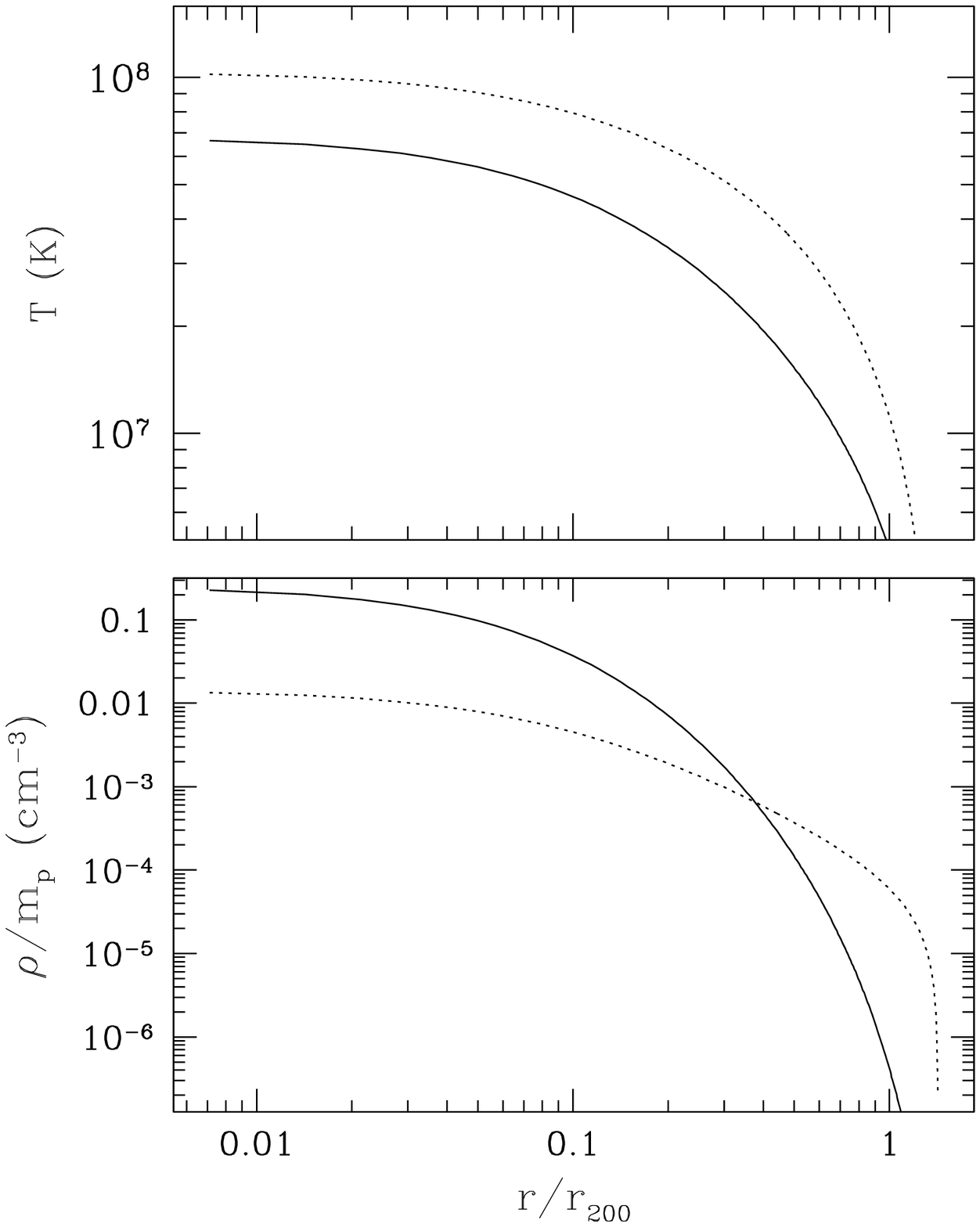}
\includegraphics[height=.36\textheight,width=.48\textwidth]{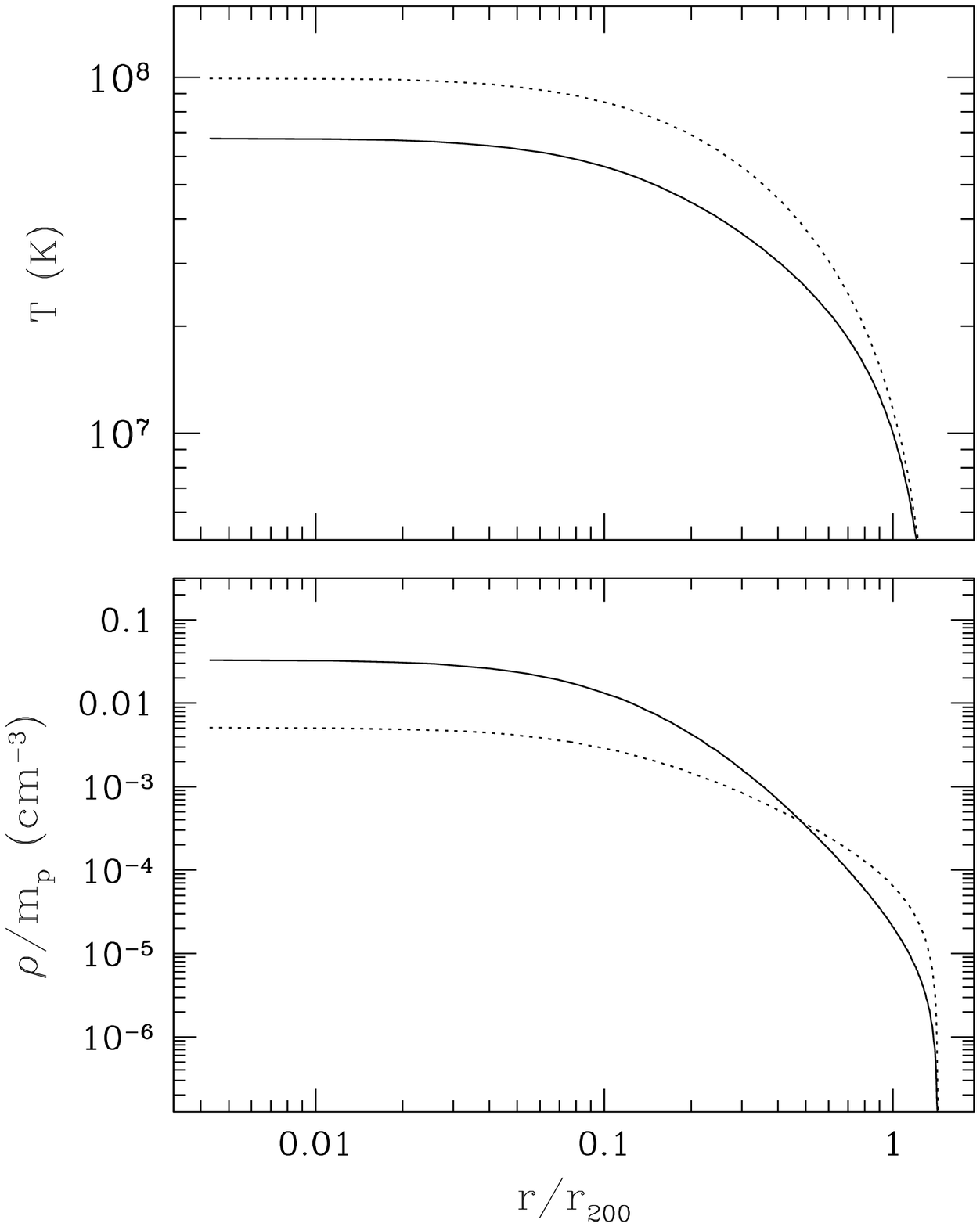}
\caption{Radial profiles of density and temperature for truncated
quasi-polytropic spherical gas models of index $\Gamma =1.2$,
in the Hernquist potential (left) and
$\gamma=0$ potential (right), for the same $\Tvir=2.3$ keV
and mass parameters
of Fig.~\ref{iso}. The central 
temperature of the parent
polytropic distribution is $\Tz=2.6\,\Tvir$ (solid lines) and
$\Tz=4.0\,\Tvir$ (dotted lines).}
\label{poly}  
\end{figure*}

A different approach, that we do not explore here (but that could be
easily implemented in our scheme), would be that of fixing the
pressure to some prescribed value on the truncation surface, by
imposing a finite density jump at $\phit$, as done in Ostriker et
al. (2005).  Note that the central values of $T$ and $\rho$ of the
truncated distribution are not $\rhoz$ and $\Tz$ of the isothermal
parent distribution in eq.~(\ref{rhois}), and the temperature at the
truncation surface vanishes.  Formally, the untruncated case (i.e.,
the true isothermal case) is recovered for $\phit\to\infty$, {\it or}
for $\Tz \to 0$. At the opposite case, i.e., for very large $\Tz$, the
following asymptotic behavior is obtained:
\[
\rho\sim\rhoz {\phit -\phi\over \betaz},\quad
T\sim {\Tz(\phit-\phi)\over 2\betaz},\quad \betaz\gg 1 .
\label{istas}
\]
In this limit the temperature distribution becomes {\it
independent} of $\Tz$, and $p\propto\rho^2$. Also the asymptotic density
profile, for an assigned gas mass, is independent of
$\Tz$.  

Summarizing, a quasi-isothermal model is determined by choosing a mass
model as described in Sect. 2.1, and by assuming $\phit = \phi
(\rvir)$ (that we arbitrarily fix along the $x$-axis, see
eqs.~[\ref{phicz}]-[\ref{phico}]).  Then a $\Tz$ is chosen and $\rhoz$
is obtained by imposing that the total $\Micm$ of the truncated
distribution equals a prescribed value.  Figure~\ref{iso} shows the
density and temperature profiles of quasi-isothermal equilibria in a
$\gamma=1$ and $\gamma=0$  spherical mass distribution.
The total dark matter mass is $M=5\times 10^{14}\Msun$ and we assume
$\Micm =0.14 M$, according with the direct measurements of gas mass
fractions of LaRoque et al. (2006), for the concordance flat
$\Lambda$CDM model. As expected, flatter temperature profiles are
obtained for lower values of $\Tz/\Tvir$, while for high values of
$\Tz/\Tvir$ the density profile tends to the limit distribution~(\ref{istas}).
In case of intermediate dark matter flattenings (e.g.,
$\epsilon=0.1$, $\eta=0.3$), the maximum flattening of the gas
distribution is $\simeq 0.10$ in the $(x,z)$ plane, while in the
$\epsilon=\eta=0.5$ case the maximum gas flattening is $\simeq 0.16$.
These figures are similar in the $\gamma=0$ and $\gamma=1$ models, and
go in the expected direction. The reason for this 
lies in the well known fact that the gas density and temperature
distributions are stratified on equipotential surfaces, that are much
less flattened than the mass distribution that produces them (e.g.,
Binney \& Tremaine 2008). Therefore, even for the flattest mass
distributions that can be allowed, the corresponding density profiles
keep roundish.

\subsubsection{Truncated quasi-polytropic models}
\label{trupol}

Polytropic models are equilibrium statifications for which $p=\pz
(\rho/\rhoz)^{\Gamma}$ and $T=\Tz (\rho/\rhoz)^{\Gamma-1}$, with the
polytropic index $1<\Gamma\leq 5/3$, and $\rhoz$ and $\Tz$ are (for
example) the central values of the gas density and temperature,
respectively. These models are more complicate than isothermal
stratifications. In fact, in this case the solution of the hydrostatic
equilibrium can be written as
\[
\left({\rho\over\rhoz}\right)^{\Gamma -1}={T\over\Tz}=
1-{\Gamma -1\over\Gamma\betaz}(\phi-\phiz), \quad \betaz\equiv {k\Tz\over
\mu \mH},
\label{rhop}
\]
where $\betaz$ now
refers to the central value of the temperature. It follows that, given
the depth of the potential well, a critical
temperature
\[
\Ttr\equiv {\Gamma-1\over\Gamma}{\mu\mH |\phiz|\over k}
\label{tcr}
\] 
exists so that for $\Tz\geq\Ttr$ the distribution in eq.~(\ref{rhop})
is untruncated, and the total gas mass diverges. For $\Tz =\Tzt <\Ttr$
instead a truncation value $\phit$ defined by the identity
\[
\Tzt ={\Gamma-1\over\Gamma}{\mu\mH (\phit -\phiz)\over k}
\label{tempz}
\]
exists, so that $T(\phit)=0$. Alternatively, having fixed the two 
values $0>\phit >\phiz$ for the potential, 
only one temperature $\Tzt$ exists that produces a
naturally truncated polytrope at the surface $\phi=\phit$.  However,
it can be useful to have a whole family of quasi-polytropic models
truncated at $\phit$ for {\it all} 
temperatures $\Tz\geq\Tzt$. This can be obtained
following the same approach as in Sect. 2.2.1. Thus, for given $\phit$
and $\Tz\geq\Tzt$, we introduce the truncated density
\[
{\rho\over\rhoz}\equiv
\left({T\over\Tz}\right)^{{1\over\Gamma-1}}-
\left({\Tt\over\Tz}\right)^{{1\over\Gamma-1}},
\label{rhopt}
\]
where $T$ is the temperature of the parent model~(\ref{rhop}), and
$\Tt$ is its value at $\phit$; of course $\Tt=0$ for $\Tz=\Tzt$.
Following the method described in Appendix A, the quasi-polytropic
equilibrium temperature corresponding to eq.~(\ref{rhopt}) is
\[
{T\over\Tz}={\displaystyle{
\left({T\over\Tz}\right)^{{\Gamma\over\Gamma-1}}-
\left({\Tt\over\Tz}\right)^{{\Gamma\over\Gamma-1}}-
{\phit -\phi\over\betaz}\left({\Tt\over\Tz}\right)^{{1\over\Gamma-1}}}
\over 
\displaystyle{\left({T\over\Tz}\right)^{{1\over\Gamma-1}}-
\left({\Tt\over\Tz}\right)^{{1\over\Gamma-1}}}},
\label{tempt}
\]
where the temperature distribution at the r.h.s. is that given by
eq.~(\ref{rhop}).  Summarizing, after having choosen a dark matter
distribution and the value $\phit=\phi(\rvir)$ as in the
quasi-isothermal case, the associated $\Tzt$ is calculated.  A
truncated quasi-polytropic model is then determined by fixing a
temperature $\Tz\geq\Tzt$, so that $\Tt$ is determined through
eq.~(15), and $\rhoz$ is obtained so that $\Micm$ of the truncated
distribution~(18) coincides with the required value.

\begin{figure*}
\hskip -0.3truecm
\includegraphics[height=.36\textheight,width=.38\textwidth]{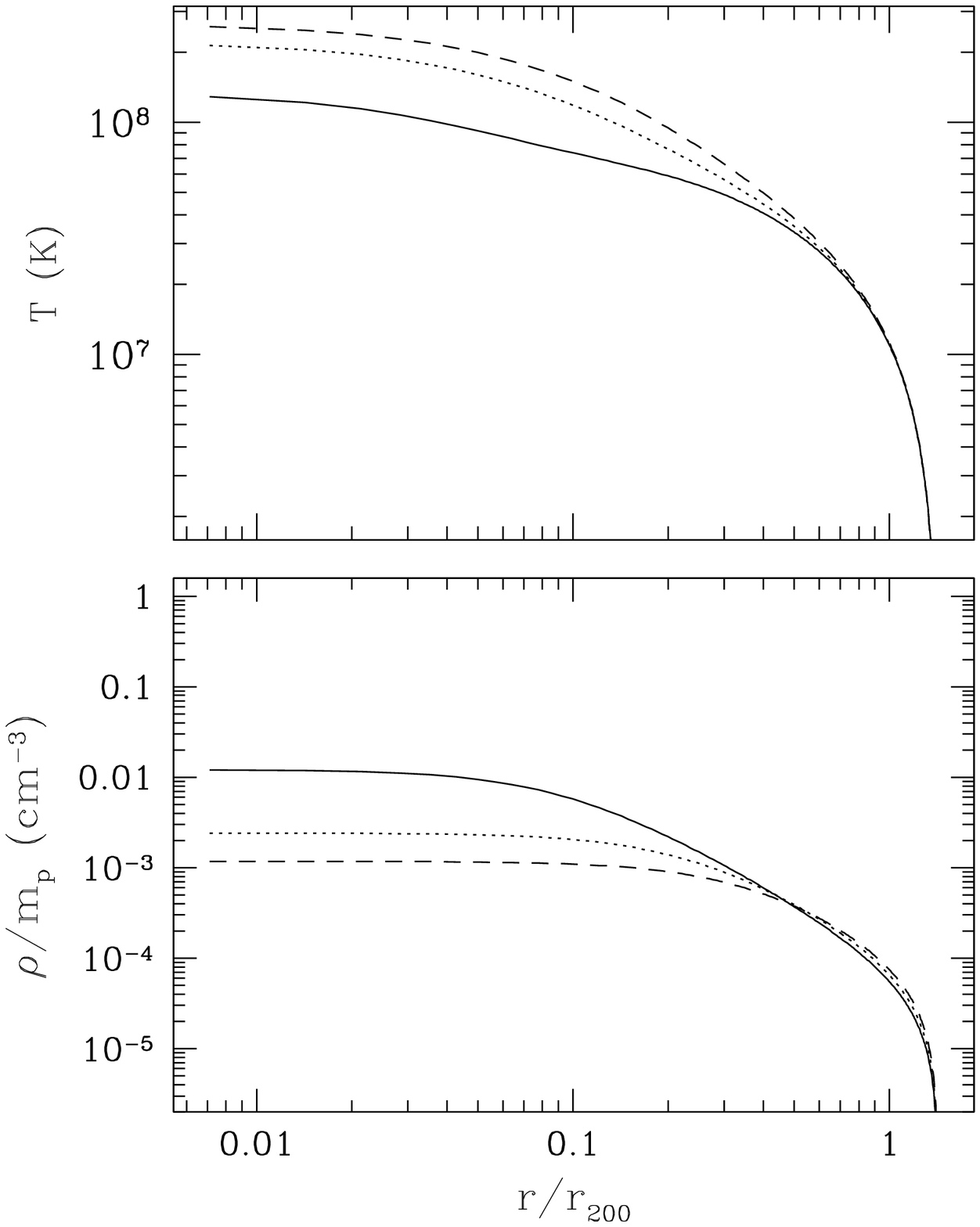}
\hskip -1truecm
\includegraphics[height=.36\textheight,width=.38\textwidth]{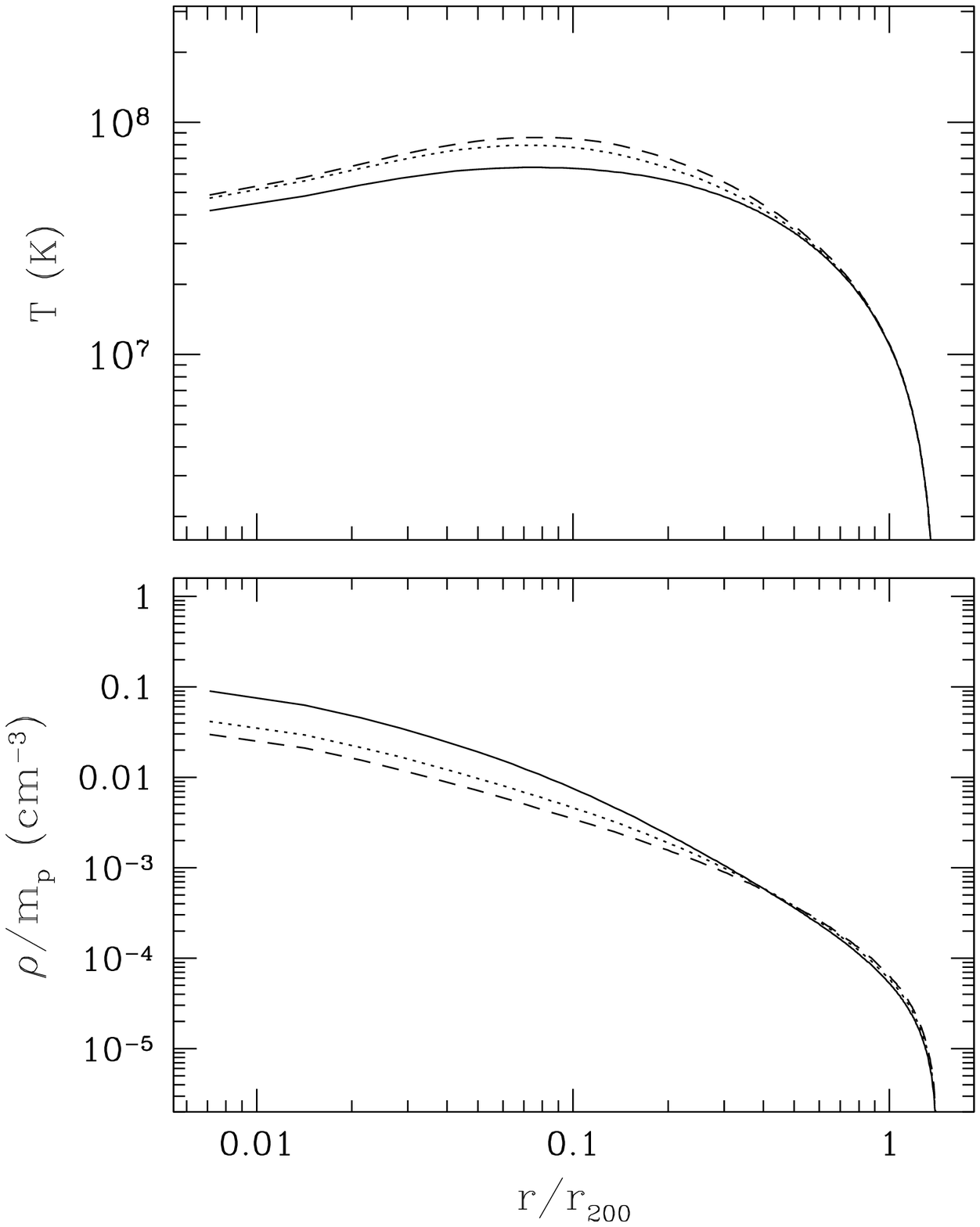}
\hskip -1.7truecmls
\includegraphics[height=.36\textheight,width=.38\textwidth]{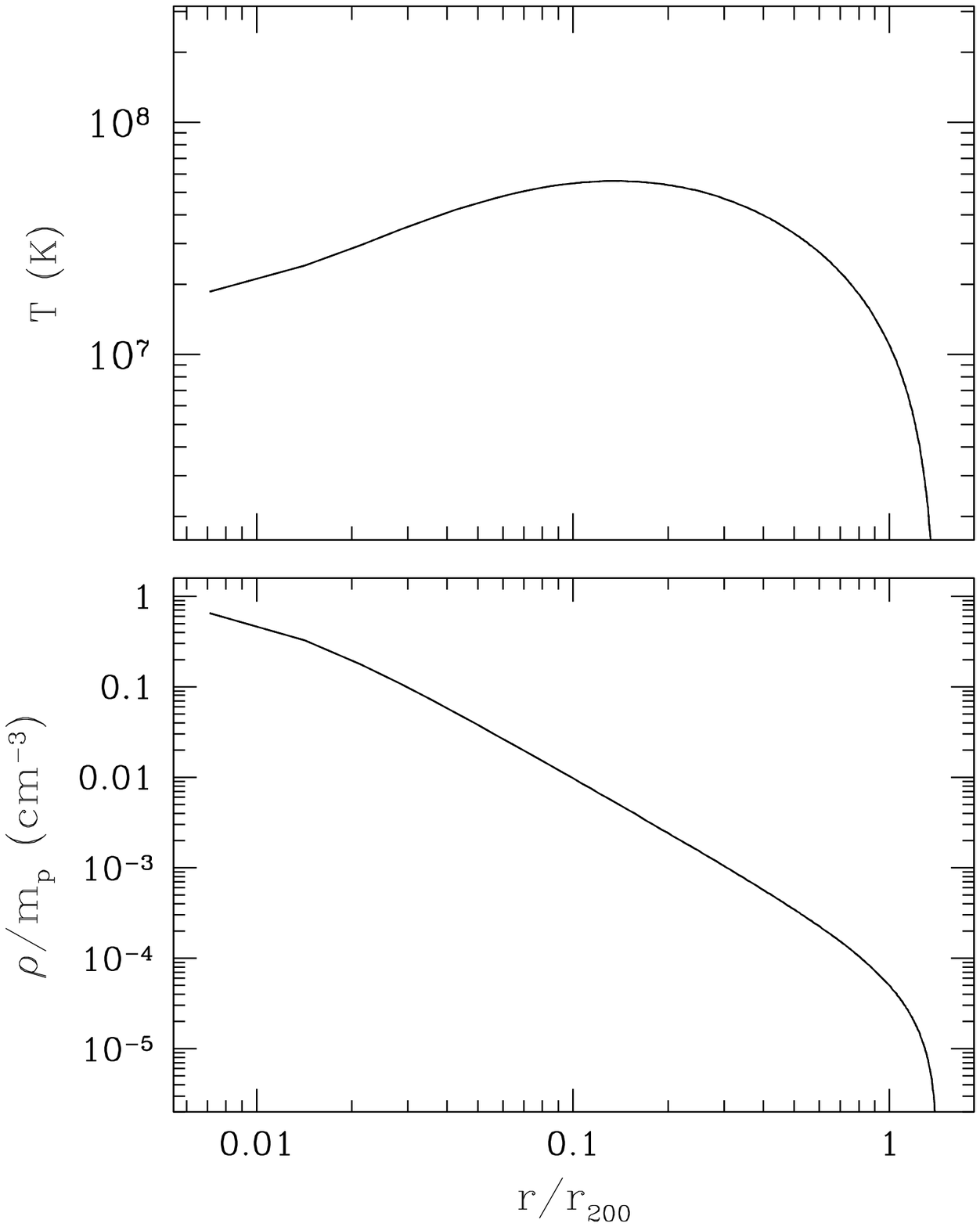}
\caption{
Gas density and temperature radial profiles for a TMB
model (with $\beta=2/3$) in equilibrium within the spherical
Hernquist potential used in Figs. 1 and 2. From
left to right the panels refer to $\alpha =0$, $\alpha =1$ and
$\alpha=2$.  In each panel the solid, dotted and dashed lines refer to
$\rg / \rc =$ 0.4, 1, and 1.6, respectively.  Note that
consistently with eq.~(\ref{rhobaro}) the profiles of the $\alpha=2$
models are independent of $\rg / \rc$.  }
\label{baro}  
\end{figure*}

We remark that the pair~(\ref{rhopt})-(\ref{tempt}) when $\Tz = \Tzt$ 
reduces to the polytrope naturally truncated at $\phit$,
while for very high values of the central temperature 
\[
\rho\sim\rhoz {\phit -\phi\over \Gamma\betaz},\quad
T\sim {\Tz(\phit-\phi)\over 2\betaz},\quad \betaz\gg 1,
\label{poltas}
\]
and, as in the quasi-isothermal case, the temperature distribution
becomes independent of $\Tz$.  For reference, from eqs.~(\ref{tempv}),
(\ref{istas}) and (20) it follows that for the limit $\gamma=1$ models
the ratio of the true central gas temperature $T(0)$ to $\Tvir$ is
$\simeq 7.7$, while in the limit $\gamma=0$ models it is $T(0)\simeq
6.2\Tvir$. 

Figure~\ref{poly} shows the density and temperature profiles for
quasi-polytropic spherical models with $\Gamma=1.2$ (a value reported
to produce a good fit of some observed temperature profiles for the
ICM, Markevitch et al. 1998) in the same potentials adopted for
Fig.~\ref{iso}.  As for the truncated quasi-isothermal models, steeper
density profiles in the central regions are obtained for the
$\gamma=1$ than for the $\gamma=0$ potential, to balance the steeper
potential well (even though in the quasi-polytropic case the
steepening can be minor, being in part compensated by the temperature
increase towards the center).  Note that models analogous to the
"coldest" quasi-isothermal models in Fig.~\ref{iso} do not exist
because from eq.~(\ref{tempz}) the minimum admissible temperature
$\Tzt$ is $2.1\Tvir$ for $\gamma=0$ and $2.6\Tvir $ for $\gamma=1$.
As in the quasi-isothermal cases, also here the effect of dark matter
flattening on the density and temperature distributions is quite
modest.  In fact, being the gas stratified on the potential, the
flattenings of the gas distributions are the same as described at the
end of Sect.~\ref{truiso}.

\subsubsection{Truncated modified $\beta$ models}
\label{barot}

The models introduced in the previous Sects.~\ref{truiso} and
\ref{trupol} are just two special barotropic families built starting
from prescribed relations $p(\rho)$; as a consequence, their density
profile is somewhat out of control. Here we show how to derive the
temperature distribution for an hydrostatic gas of assigned density
profile in an external potential well deviating from spherical
symmetry.  We call this approach "density approach"\footnote{For the
more complicate case of the construction of rotating, baroclinic
gaseous distributions, see Barnab\`e et al. (2005).} and technical
details are given in Appendix A2. In practice, the idea behind the
method is to construct the spherical barotropic solution for a given
gas density profile in a given spherical potential, and then to {\it
deform} (maintaining the equilibrium) the potential and the gas
density distribution: this is accomplished by constructing the
integral function $H$.

As relevant case for the present discussion, the starting density
distribution is a spherical truncated modified $\beta$--model (hereafter
TMB)
\[
{\rho\over\rhoz}=\left({\rg\over r}\right)^{\alpha} 
                 \left(1+{r^2\over\rg^2}\right)^{\alpha-3\beta\over 2}-
                 \left({\rg\over\rt}\right)^{\alpha} 
                 \left(1+{\rt^2\over\rg^2}\right)^{\alpha-3\beta\over 2}
\label{rhobaro}
\]
for $r\leq\rt$, with $\rg$ a core radius and $\rt$ a truncation
radius. This density profile is a modification of the well known
$\beta$-model (Cavaliere \& Fusco Femiano 1976) and its generalization
(Lewis et al. 2003).  In particular, the density is proportional to
$r^{-\alpha}$ for $r\to 0$, and to $r^{-3\beta}$ for $\rg\ll r\ll\rt$.
A finite gas mass is obtained for $0\leq\alpha <3$, and for $\beta>1$
no truncation would be required. Here the introduction of $\rt$ is
needed because $0.5\lsim \beta\lsim 0.8$ from fits to observed ICM
profiles (e.g., Mohr et al. 1999, Jones \& Forman 1999).  For a
spherical Hernquist potential, the density approach applied to
eq.~(\ref{rhobaro}) leads to the function
\begin{eqnarray}
{\rho(\Psi)\over\rhoz b^{3\beta}}&=&{\Psi^{3\beta}\over
             (1-\Psi)^{\alpha}[(1+b^2)\Psi^2-2\Psi+1]^{(3\beta-\alpha)/2}}\cr
       &&-{\Psit^{3\beta}\over
         (1-\Psit)^{\alpha}[(1+b^2)\Psit^2-2\Psit+1]^{(3\beta-\alpha)/2}},
\label{rhobart}
\end{eqnarray}
where $\Psi\equiv \phi/\phiz$ is the Hernquist potential normalized to
its central value, and $b\equiv\rg /\rc$.  Note how the two limiting
cases of very small and very large $b$ correspond to truncated power
law gas distributions: $\rho\propto r^{-3\beta}- \rt^{-3\beta}$ for
$b\to 0$, and $\rho\propto r^{-\alpha} -\rt^{-\alpha}$ for
$b\to\infty$.  The function $H(\Psi)$ needed to determine the
temperature distribution (eqs.~[A4], [A6]) cannot be expressed in terms
of elementary functions for generic values of $\alpha$ and $\beta$;
however, simple cases are obtained for $\alpha=0,1,2$ and
$\beta=(\alpha+n)/3$ with $n$ non-negative integer.  The explicit
formulae for $\alpha=0,1,2$ and $\beta=2/3$ (that falls within the observed
range quoted above) are provided in Appendix
A2, and hereafter only these values will be used. 
Thus small values
of $\rg/\rc$ correspond to models converging to the truncated $r^{-2}$
profile, independently of the specific value of $\alpha$, while for
$\alpha=2$ the distribution is independent of $b$.   The final
step of the procedure is to substitute the deformed potential given in 
eq.~(\ref{phico}) in eq.~(\ref{rhobart}) and in the function $H$,
since by construction all the resulting formulae are still exact when the
potential is deformed to the axisymmetric or triaxial case.

Figure~\ref{baro} shows the density and temperature profiles for the
$\alpha=0,1,2$ spherical cases.  The temperature decline in the
$\alpha=1$ and $\alpha=2$ models compensates the steep increase of
$\rho$, in order to produce the pressure gradient needed to
balance the imposed gravitational field.  In a broad sense, this
behavior is similar to that of the velocity dispersion profile in the
central regions of isotropic Hernquist or $R^{1/m}$ models (Ciotti
1991). For $\alpha =1$, lower values of $b$ correspond to a more
important central peak of the density profile and a more important
decline of the temperature in the central region.  Thus, although the
central temperature drop is not due to cooling, these models
provide an interesting {\it phenomenological} description of cool-core
systems. The opposite behavior is shown by the $\alpha=0$ models, in
which the flat-core gas density requires central temperatures higher
than in all the other cases.  Finally, the introduction of flattening in the
dark matter halos does not lead to significant deformations in the gas
density distributions, with maxium deviations as reported at the end
of Sects.~\ref{truiso} and \ref{trupol}.

\begin{figure*}
\vskip -1truecm
\hskip -1.1truecm
\includegraphics[height=.46\textheight,width=.58\textwidth]{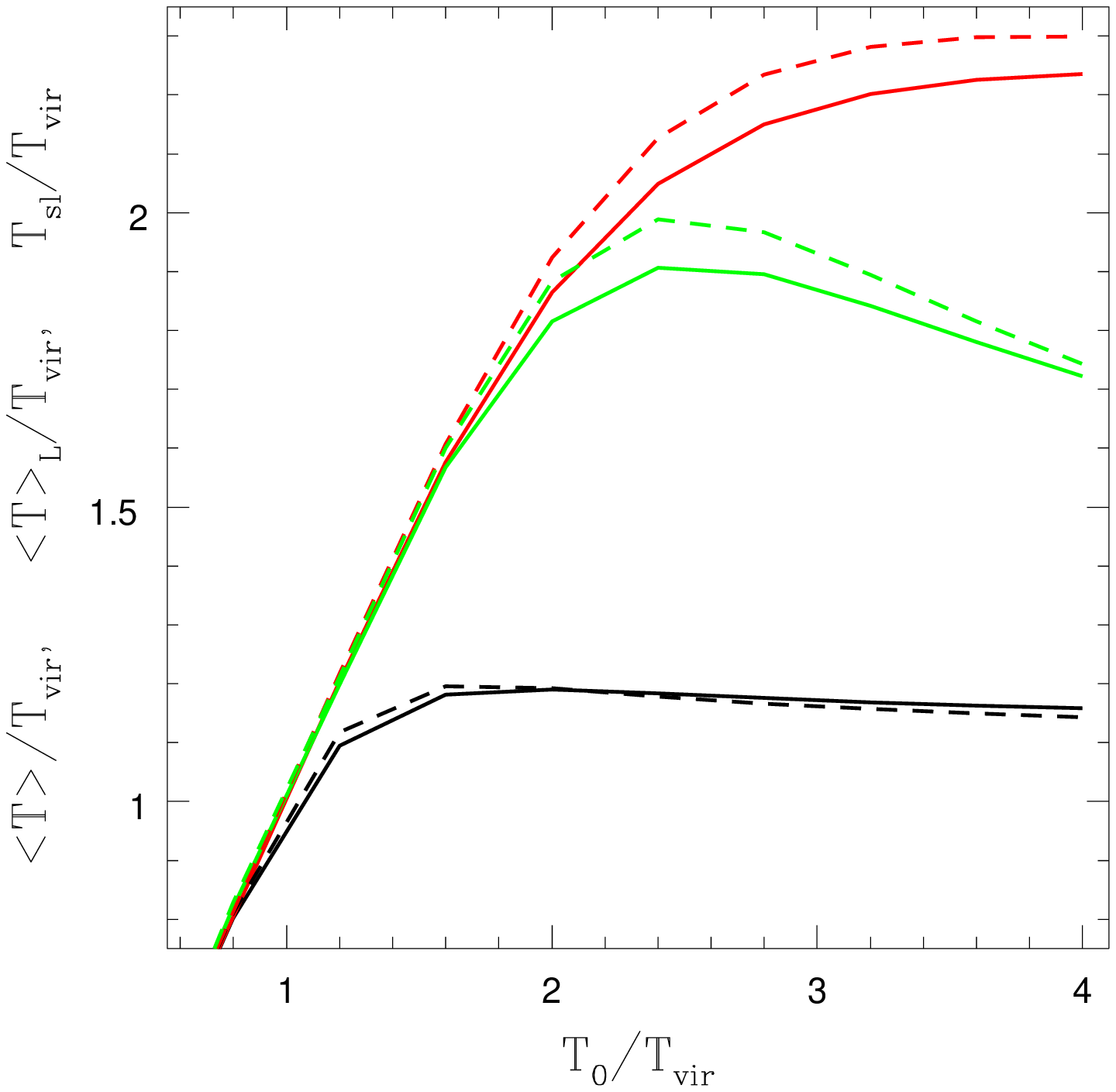}
\hskip -1.9truecm
\includegraphics[height=.46\textheight,width=.58\textwidth]{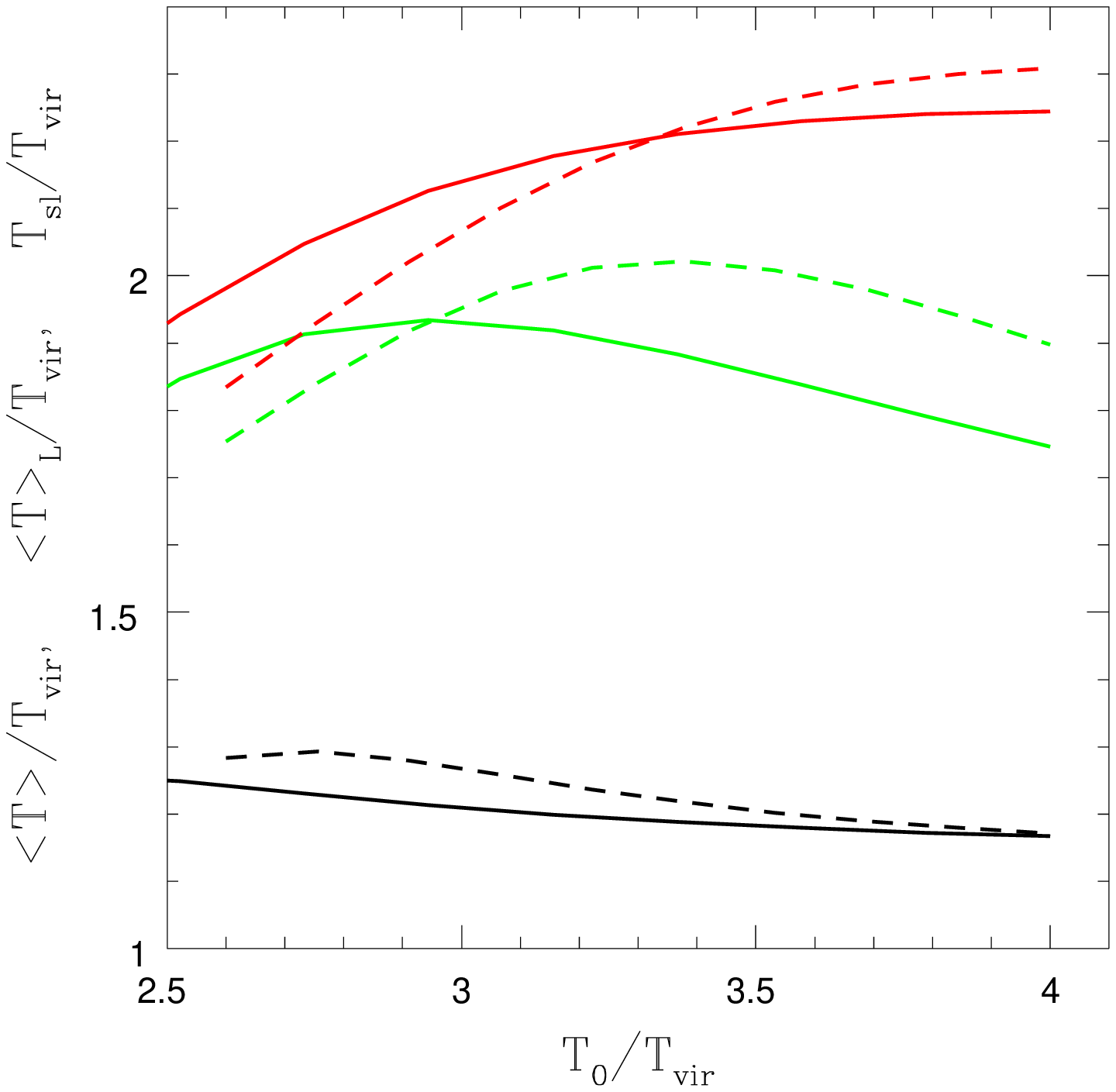}
\caption{
The mass weighted $\TicmM$ (black), luminosity weighted $\TicmL$ (red)
and spectroscopic-like $\Tspec $ (green) temperatures
for the truncated quasi-isothermal models (left) and the truncated
quasi-polytropic models with $\Gamma=1.2$ (right).  Solid lines refer
to the $\gamma=0$ potential, dashed lines to the Hernquist potential
(see Sect.~\ref{tempav} for more details).}
\label{tempiso}  
\end{figure*}

\subsubsection{Comparison with observed ICM properties}\label{comp}

Even though the aim of this work is not to construct models
reproducing in detail the observed ICM properties (which is hard
within the simple framework of hydrostatic equilibrium of single-phase
gas in smooth potential wells), we briefly comment here on how the
obtained equilibria compare with observations.  In general, the
quasi-isothermal and polytropic models, and the TMB models with
$\alpha=0$, are similar to ``non cool-core'' systems, while the
$\alpha=1$ and $\alpha =2$ TMB models, where the temperature profile
is decreasing towards the center, are similar to ``cool-core''
systems.

In the family of non cool-core models, quasi-isothermal distributions
can be built with arbitrarily low temperatures, becoming more and more
similar to the standard isothermal models. Quasi-polytropic models
instead, once the truncation potential is fixed, cannot be built with
a central temperature smaller than a limit temperature roughly
corresponding to the depth of the dark matter potential well.  In the
past, polytropic models with $\Gamma=1.2$ have been used to reproduce
the external regions of ICM observed (e.g., Markevitch et al. 1998,
Piffaretti et al. 2005) and simulated (Ostriker et al. 2005).

In the family of cool-core models, TMB distributions with $\alpha=1$
or $\alpha=2$ show temperature profiles in good agreement with those
observed by $Chandra$ and $XMM-Newton$ (e.g., Allen et al. 2001,
Kaastra et al. 2004, Vikhlinin et al. 2005, 2006); on average, these
profiles reach a maximum near $\rc$ and then decline at larger radii,
reaching $\sim 0.5$ of their peak value near $r\sim 0.5 \rvir$. In
addition, not only the profile shapes of these TMB models are similar
to the observed ones, but also their temperature values, when rescaled
to the mass-weighted temperature within $r_{500}$ ($T_{500}$), agree
with the observed values (as those shown by Viklinin et al. 2006).
The relation between $\TicmM$ and $T_{500}$ will be briefly addressed
at the end of Sect. 3.1.

\subsection{The abundance profile and the emissivity}
\label{emuff}

In addition to the dark matter potential well and the hydrostatic gas
distribution, the third ingredient of our models is the metal
distribution.  In the numerical code the metal abundance profile is
assumed to be stratified according to a formula which generalizes to
the ellipsoidal case the observed abundance profiles (Ikebe et al.
2003, De Grandi\footnote{The De Grandi et al. (2004)'s formula does
  not have the square of the radial coordinate, but we found that the
  square is needed to match the datapoints in their Fig. 2.} et
al. 2004, Vikhlinin et al. 2005), i.e.
\[
Z = {Z_0 \over (1+\mz^2)^{\zeta}}, \quad 
\mz^2 = {x^2 \over \rz^2} + {y^2 \over (1-\epsilon)^2\rz^2}
      + {z^2 \over (1-\eta)^2\rz^2},
\label{degra}
\]
where the central metallicity is $Z_0=0.8\Zsun$, the slope
$\zeta=0.18$ and the metallicity scale-length $\rz=0.04\rvir$. In
addition, the flattening of the metallicity distribution is the same
used for the dark matter distribution. Obviously, we are not attaching
any special physical reason to this last assumption, except to have
flatter metal distributions in flatter systems, and to reduce the
parameter space dimensionality. In any case, we also explored cases
where the metals are stratified exactly on
isodensity surfaces, i.e. $Z\propto (\rho/\rhoz)^{\zeta}$, without
finding significant differences with the case of eq.~(\ref{degra}).

\begin{figure}
\vskip -2truecm
\hskip -0.4truecm
\includegraphics[height=.46\textheight,width=.61\textwidth]{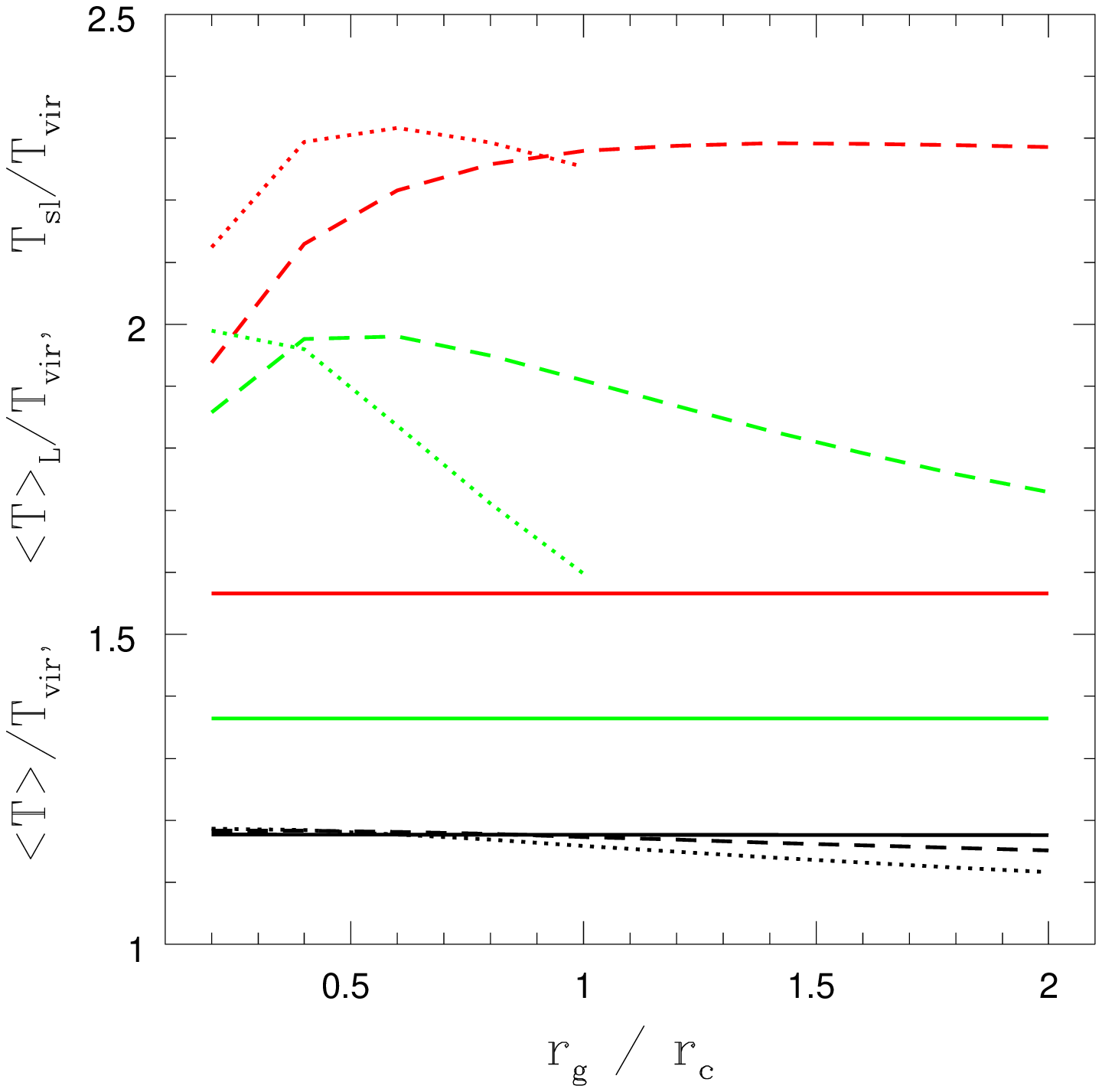}
\caption{
The mass weighted $\TicmM$ (black), luminosity weighted $\TicmL$ (red)
and spectroscopic-like $\Tspec$ (eq.~[\ref{tvik}], green)
temperatures calculated for TMB gas models
within the $\gamma=1$ potential, for $\alpha =0$ (dotted lines),
$\alpha=1$ (dashed lines) and $\alpha=2$ (solid lines), as a function
of $\rg/\rc$.  For $\alpha=2$ the average temperatures are independent
of $\rg/\rc$, consistently with eq.~(\ref{rhobaro}) with $\beta=2/3$.
For $\alpha=0$ the value of $\TicmL/\Tvir$ is displayed only up to $\rg/\rc=1$
since for higher $\rg/\rc$ the central temperatures
become very high (Fig.~\ref{baro}).}
\label{tempbar}  
\end{figure}


The emissivity adopted in the code is given by 
\[
\Emis = n_e n_H\Lambda (T, Z)\quad 
\label{emiss}
\]
where $n_e$ and $n_H$ are the number densities of electrons and
hydrogen. The cooling function $\Lambda (T,Z)$ has been calculated
over the energy interval 0.3--8 keV with the radiative emission code
APEC for hot plasmas at the collisional ionization equilibrium (Smith
et al. 2001), as available in the XSPEC package (version 12.2.0) for
the solar abundance ratios of Grevesse \& Sauval (1998).  With APEC we
have computed a matrix of values for $\Lambda(T,Z)$ for a very large
set of temperatures and metallicities.  Note that the cooling function
can be written as
\[
\Lambda(T,Z)=\Lambda(T,0)\left[1+Z\,g(T,Z)\right],
\label{eqlambda}
\]
where $Z$ is in solar units, $\Lambda (T,0)$ is the function in the case 
of no metals, and 
\[
g(T,Z) \equiv {\Lambda(T,Z)-\Lambda(T,0) \over Z\, \Lambda(T,0)}.
\]
It turned out that the function $g$ is almost exactly independent of
$Z$, so that eq.~(\ref{eqlambda}) with $g=g(T)$ exploits the nearly
perfect linear dependence of the function $\Lambda(T,Z)$ on abundance.
In order to speed up the numerical code we computed non-linear fits of
the functions $\Lambda(T,0)$ and $g(T)$ valid over the temperature
range 0.1--16 keV (with maximum deviations from the APEC values
$<1$\%) and reported in Appendix B. We remark that $g(T)$ declines
steadily with increasing $T$, from $\simeq 42$ down to $\simeq 0.14$,
which has the consequence that for {\it high} values of $Z
g(T)$ then $\Lambda(T,Z)\propto Z_0$ and $\Lx\propto Z_0$, while
for {\it low} values of $Z g(T)$ both $\Lambda(T,Z)$ and $\Lx$ are
independent of $Z_0$.

\begin{figure*}
\vskip -2truecm
\hskip -0.9truecm
\includegraphics[height=.46\textheight,width=.55\textwidth]{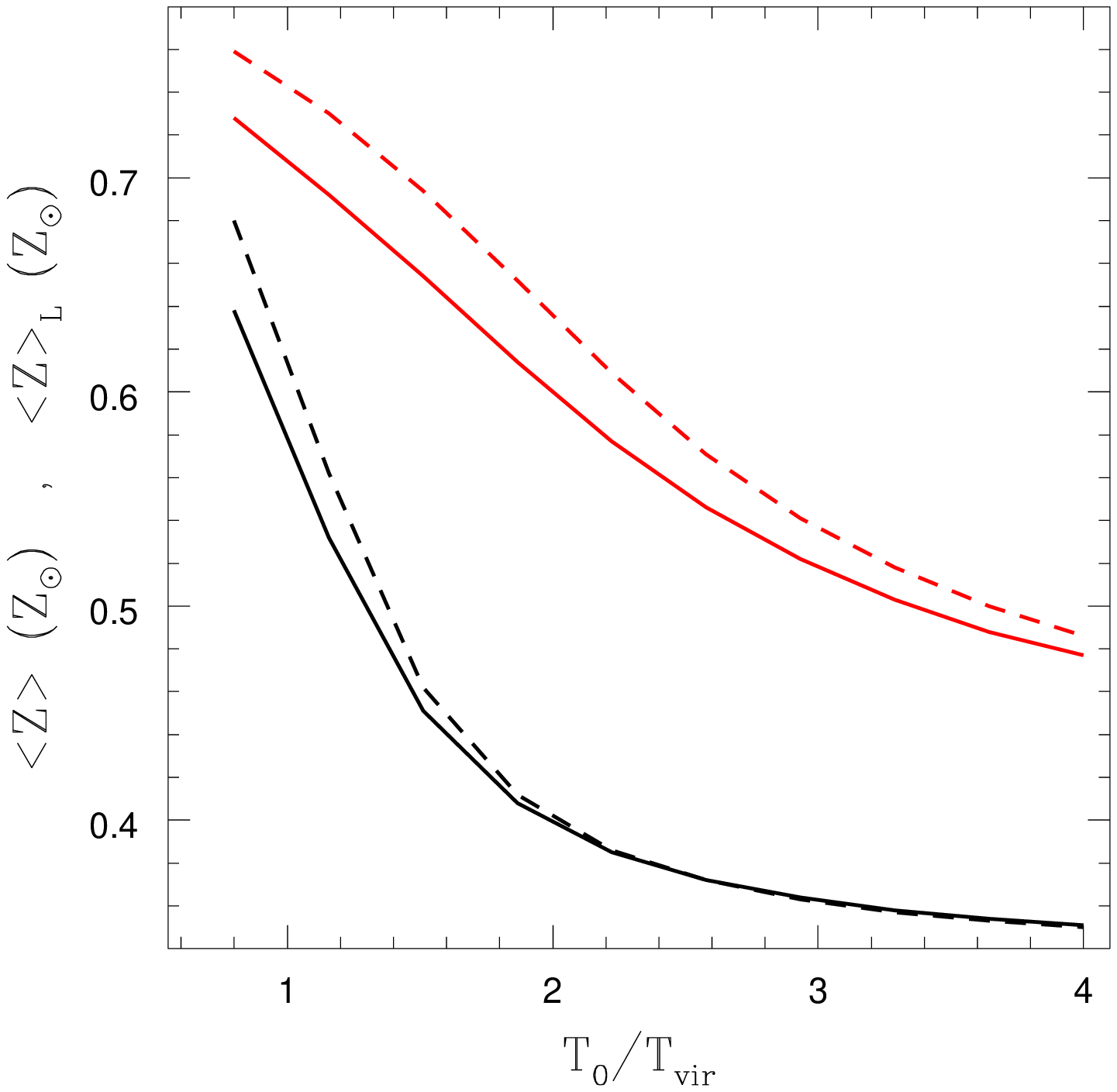}
\hskip -2truecm
\includegraphics[height=.46\textheight,width=.55\textwidth]{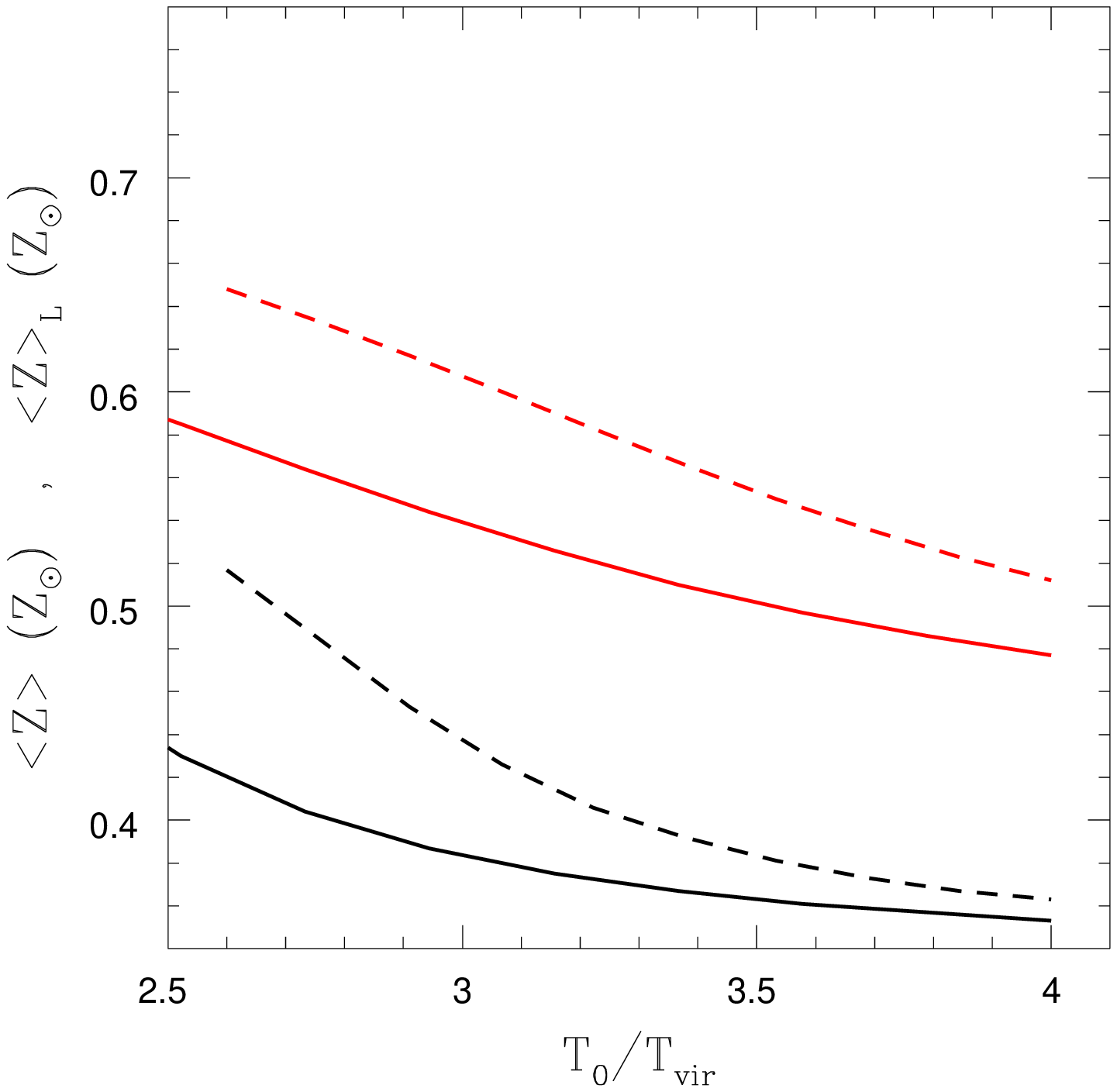}
\caption{The mass weighted $\ZicmM$ (black) and luminosity weighted
  $\ZicmL$ (red), calculated for quasi-isothermal equilibria (left)
  and for quasi-polytropic equilibria with $\Gamma =1.2$ (right),
  within the $\gamma=0$ (solid lines) and $\gamma=1$ (dashed lines)
  potentials.  The abundance profile is that of eq.~(\ref{degra}),
  with $Z_0=0.8\Zsun$.}
\label{metiso}  
\end{figure*}

\section{Results}\label{res}

For each model the quantities $\ZicmL$ and $\TicmL$ are not computed
through projection, but directly as volume integrals. In fact, from
the Projection Theorem (Appendix A1) eqs.~(4)-(5) can be also written
as
\[ 
\ZicmL =
\displaystyle{\int\Emis [\rho(\xv),T(\xv),\Zicm (\xv)] \, \Zicm
(\xv) \, d^3\xv\over \Lx}, 
\label{eqzicml} 
\]
and
\[
\TicmL = 
\displaystyle{\int\Emis  [\rho(\xv),T(\xv),\Zicm (\xv)] \, T (\xv) \, 
d^3\xv\over \Lx},
\label{eqticml} 
\]
where $\Emis $ is the emissivity in the 0.3--8 keV band due to gas in
the temperature range 0.1-16 keV.  As anticipated in the Introduction,
for each model we also compute the spectroscopic-like temperature.  
Following  Vikhlinin (2006), this is estimated as
\[
\Tspec=xT_{\rm cont}+(1-x)T_{\rm line},
\label{tvik}
\] 
where $T_{\rm cont}$ and $T_{\rm line}$ are the continuum-based and
line-based temperatures for the composite spectrum and $x$ is a
parameter that measures the relative contribution of the line and
continuum emission to the total flux.  To evaluate $T_{\rm cont}$,
$T_{\rm line}$ and $x$, three functions of the temperature are needed;
these depend on the instrument in use, energy band, redshift and
neutral hydrogen absorbing column. As a representative case we chose
to simulate observations made with the $Chandra$ CCDs over the 0.3--8
keV band, for a plasma at zero redshift and zero absorbing column.
A. Vikhlinin kindly provided us with the tabulated values of the
required functions, that we fitted with the same high precision method
described in Appendix B and we then inserted in our code.  We recall
that the method of Vikhlinin (2006) holds for thermal components of
$kT\gsim 0.5$ keV.  We also computed the estimate of $\Tspec$ proposed
by Mazzotta et al. (2004) for plasma components at $kT\gsim 3$ keV:
\[ 
\Tspec (\delta)={\int \rho^2 (\xv) T^{\delta-1/2} (\xv) d^3\xv\over 
        \int \rho^2 (\xv) T^{\delta-3/2} (\xv) d^3\xv}, 
\label{tspec}
\] 
where $\delta=3/4$ for observations obtained with $Chandra$ and
$XMM-Newton$.

It is important to note that if $\Zicm $ and $T$ do not depend on
$\xv$, then\footnote{ This identity in the case of $\Tspec$ of
  Vikhlinin (2006) can be easily proven considering the explicit form
  of the functions entering in eq.~(\ref{tvik}).}  $\ZicmL =\ZicmM
=\Zicm $ and $\TicmL=\TicmM=\Tspec=T$, even for density distributions
depending on $\xv$; otherwise the variously weighted quantities
differ, in a way dependent on the spatial distribution of $\rho$, $T$
and $\Zicm$. A quantitative estimate of these differences is the task
of the following Sects.~\ref{tempav} and~\ref{zav}.

For each model all integrals have been calculated numerically with a
double-precision code. The integration scheme employs a linear
interpolation of the gridpoint-defined variables, and the number of
grid points in the positive octant of the $(x\,,y\,,z)$ space is
$n_x\times n_y\times n_z= 300\times 300\times 300$.  Checks of the
code have been performed by calculating (with both linearly spaced and
logarithmic grids) the total masses of strongly peaked triaxial
distributions whose values are known analytically, and also mean value
temperatures of special distributions for which the expected values
can be calculated analytically (see Section 3.1), obtaining errors
$\lsim$ 0.1\%.

\subsection{Temperature averages}
\label{tempav}

Figures~\ref{tempiso} and \ref{tempbar} show the trend of the mass
weighted temperature $\TicmM$, of the luminosity weighted temperature
$\TicmL$ and of $\Tspec $ (eq.~[\ref{tvik}]), as a function of
$\Tz/\Tvir$ in the quasi-isothermal and quasi-polytropic cases, and of
$\rg/\rc$ in the TMB case. As for Figs.~\ref{iso}-\ref{baro},
the gravitating mass is a spherical $\gamma=0$ or $\gamma=1$ model
with $M=5\times 10^{14}\Msun$; also the range of $\Tz/\Tvir$ and
$\rg/\rc$ is the same used for Figs.~\ref{iso}-\ref{baro}.

A first general result is that at this mass $M$ the two $\Tspec$
estimates of eqs.~(\ref{tvik}) and~(\ref{tspec}) agree within $\sim
10$\% for all the explored models. The reasons for this are the
relatively flat shape of the temperature profiles that are obtained by
hydrostatic equilibria in smooth potential wells; the not too peaked
metallicity distribution; and finally the virial temperature of the
gas ($\Tvir=2.3$ keV) that is not much lower than 3 keV (i.e., the
declared limit of applicability of the Mazzotta et al.'s $\Tspec$).  A
similar finding has been reported by Rasia et al. (2008).  In fact,
for models with strongly peaked metallicity distributions, or with
much lower mass (e.g., $M=10^{14} M_{\odot}$ and $\Tvir=0.8$ keV) we
found that the two spectroscopic temperatures are clearly different,
and the Mazzotta et al. (2004) estimate would be higher than that of
Vikhlinin (2006) (up to 20\% in the explored $\Tz/\Tvir$ range, for
the quoted mass $M$).

Finally, it is useful to mention the relation between $\TicmM$ and
$T_{500}$ (see end of Sect.~\ref{comp}), since the temperature profile
is generally recovered from observations out to radii smaller than
$\rvir$, typically out to $r_{500}$ with the most sensitive
observations (e.g., Viklinin et al. 2006). The calculation of the
$T_{500}/\TicmM$ ratio for all our models confirmed the expectation
that the hotter is the central region with respect to the outer one,
the higher is this ratio. In fact, we found for quasi-isothermal
models $T_{500}=(1 \div 1.5)\TicmM$, as $\Tz/\Tvir$ goes from 0.4 to
4; for quasi-polytropic models $T_{500}=(1 \div 1.4)\TicmM$, as
$\Tz/\Tvir$ goes from 2.6 to 4.  For TMB models, $T_{500}/\TicmM$
varies respectively between 1.3 and 1.5, and between 1.3 and 1.6, for
$\alpha=1$ and $\alpha=0$, as $\rg/\rc$ varies between 0.2 to 2; for
$\alpha=2$, $T_{500}/\TicmM=1.3$, the lowest value, a consequence of
its cold central region.

\subsubsection{Quasi-isothermal and polytropic models}

We first discuss Fig.~\ref{tempiso}.  For $\Tz\gsim\Tvir$, $\TicmL$
and $\Tspec$ are both higher (up to a factor of $\sim 2$) than the
mass-weighted temperature $\TicmM$, because they are dominated by the
hotter central regions.  For quasi-isothermal models with
$\Tz\lsim\Tvir$, instead, the 3 temperatures almost coincide, because
the gas is nearly isothermal (see Fig.~1).  Overall, $\Tspec$ and
$\TicmL$ agree very well up to $\Tz\simeq 2\Tvir$.  Starting from
$\Tz\gsim 2\Tvir$, $\TicmL$ is larger than $\Tspec$, a tendency that
becomes stronger with increasing $\Tz/\Tvir$: this is due to fact that
$\Tspec$ is biased towards the lower values of the range of
temperatures [e.g. $\Tspec (3/4)$ weights each thermal component by
$\rho^2 T^{-3/4}$ instead of by $\rho^2\Lambda (T)\sim \rho^2
T^{1/2}$].

At high $\Tz$ the size of the discrepancy between $\TicmM$ and
$\TicmL$ or $\Tspec$ compares well with the analytical predictions
based on the asymptotic profiles~(\ref{istas}), (\ref{poltas}). 
From these expressions, defining $\Psi\equiv\phi/\phiz$, the
limit values of $\TicmM$ and $\Tspec (\delta)$, both in the
quasi-isothermal and quasi-polytropic cases, are
\[
\TicmM={|\phiz|\mu\mH\over2k}{\int(\Psi-\Psit)^2 d^3\xv\over
                             \int(\Psi-\Psit) d^3\xv},
\label{tmlim}
\]
and
\[
\Tspec (\delta)={|\phiz|\mu\mH\over2k}
                {\int(\Psi-\Psit)^{3/2+\delta} d^3\xv\over
                 \int(\Psi-\Psit)^{1/2+\delta} d^3\xv}, \label{tslim}
\] 
which are {\it independent of} $\Tz$.  In particular, note that
$\Tspec(2)$ corresponds to the case of pure bremsstrahlung emission,
which is similar to the emission described by our adopted cooling
function, at least for high temperatures\footnote{In the numerical
  computations we use the full expression for $\Lambda(T,Z)$ in the
  0.3--8 keV band, and this is not well represented by a simple
  power-law (see Appendix B).}.

The analytical solution of the integrals (\ref{tmlim}) and
(\ref{tslim}), for spherical $\gamma=1$ models with truncation at the virial
radius, gives $\TicmM\simeq 1.11\Tvir$, $\Tspec (2)\simeq 2.29\Tvir$
and $\Tspec(3/4)\simeq 1.30\Tvir$. These values are close to those
shown by $\TicmM$ and $\TicmL$ at $\Tz\simeq 4\Tvir$
(Fig.~\ref{tempiso}); $\Tspec(3/4)$ instead is still far from its
limit value, even though it has already started decreasing.  As long
as $\TicmL$ can be considered similar to $\Tspec(2)$, then $\Tspec
(3/4)$ is predicted to tend to $\simeq 0.6\TicmL$ (which was
verified with numerical models not presented here).
The same calculations at the limit of high $\Tz$ for the $\gamma=0$
potential give $\TicmM\simeq 1.13\Tvir$, $\Tspec (2)\simeq 2.27\Tvir$
and $\Tspec (3/4)\simeq 1.31\Tvir$; these values are very similar to
those for $\gamma=1$, in agreement with the close location of
the solid and dashed lines in Fig.~\ref{tempiso}.


\subsubsection{TMB models}

The TMB cases (Fig.~\ref{tempbar}) are more varied.  The first result
is that $\TicmM$ (black lines) remains almost constant for different
$\rg/\rc$, at a value nearly independent of $\alpha$. Analytic
integration for $\alpha =2$ shows that $\TicmM=1.18\Tvir$, in 
agreement with the numerical result in Fig.~5.  The second result is
that again $\Tspec$ and $\TicmL$ overestimate $\TicmM$.

$\Tspec$ (green lines) decreases steeply with increasing $\rg/\rc$ for
$\alpha=0$, because the density in the central hotter regions
decreases by almost an order of magnitude, while it is increasing in
the colder external region.  This same trend is again present, though
milder, for the models with $\alpha=1$.  As discussed in Appendix A2,
without a cut at low temperatures $\Tspec (\delta)$ diverges for the
$\alpha=2$ models; therefore a cut at $kT=0.1$ keV (which excludes the
cold central regions where $\rho\propto r^{-2}$) has been adopted to
produce Fig.~\ref{tempbar}.  $\TicmL$ does not suffer from this
problem, because of the low temperature cut in the cooling function.
$\TicmL$ is higher than $\Tspec$ (for the same $\alpha$) as in the
quasi-isothermal and quasi-polytropic models, and for the same reason
of being $\Tspec$ biased towards the lower values of the temperature
range.

\subsubsection{Changing the dark mass amount and shape}

For all models (quasi-isothermal, polytropic and TMB) we investigated
the effects of changing the total mass $M$ and of flattening the dark
mass distribution.  We found that all the {\it trends} in
Figs.~\ref{tempiso} and~\ref{tempbar} remain the same with a different
total mass $M$ and shape.

For what is concerning the {\it values} of the average temperatures,
$\TicmL/\Tvir$ and $\Tspec/\Tvir$ remain the same within few ($\sim
5$) percent, for $M\geq 3\times 10^{14}M_{\odot}$. The only exception
is $\Tspec/\Tvir$ calculated according to Vikhlinin (2006) for the TMB
models with $\alpha=2$, that decreases by 13\% going from $M=1.4\times
10^{15} M_{\odot}$ to $M=3\times 10^{14} M_{\odot}$.  $\TicmM/\Tvir$
and $\Tspec(\delta)/\Tvir$, instead, are independent of $M$ for all
models, as can be proved analytically: the curves in
Figs.~\ref{tempiso} and~\ref{tempbar} depend (for all the other
parameters fixed) only on $\Tz/\Tvir$ or $\rg/\rc$, and on $\epsilon$
and $\eta$.

For a fixed mass $M$ ($\geq 3\times 10^{14} M_{\odot}$), we then
changed the values of $\epsilon$ and $\eta$ from zero to 0.1 and 0.3,
which produces a flat E7-like shape, and up to 0.5 and 0.5.  The
values of all the average temperatures, when rescaled for the
different $\Tvir$, remain the same within 5\%. Again the largest
variation is that of $\Tspec/\Tvir $ calculated according to Vikhlinin
(2006) for the TMB models with $\alpha=2$, that increases by 13\%
going from the spherical to the $\epsilon=\eta=0.5$ shape (that
corresponds to a very prolate ellipsoid).

\begin{figure}
\vskip -2truecm
\hskip -0.9truecm
\includegraphics[height=.46\textheight,width=.61\textwidth]{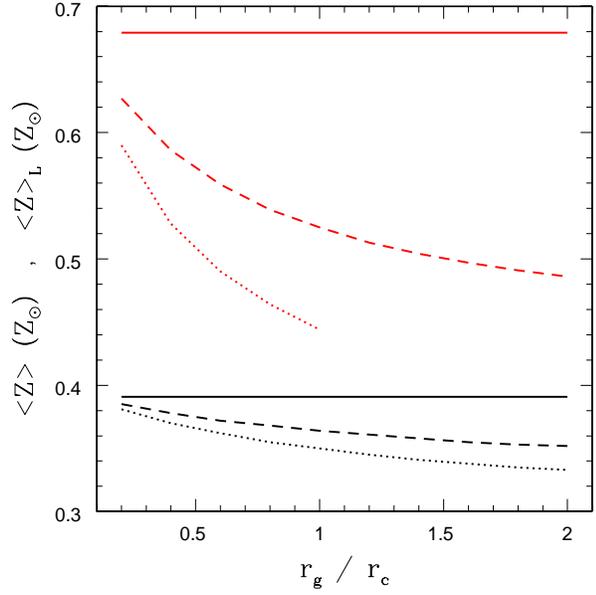}
\caption{
The mass weighted $\ZicmM$ (black) and luminosity weighted
$\ZicmL$ (red) abundances calculated for TMB models
within the $\gamma=1$ potential, for $\alpha =0$
(dotted lines), $\alpha=1$ (dashed lines) and $\alpha=2$
(solid lines), as a function of $r_g/r_c$.  For $\alpha=2$ the density
and temperature profiles are independent of $r_g/\rc$, and so are
$\ZicmM$ and $\ZicmL$; for $\alpha=0$, only the values up to $\rg/\rc=1$
are displayed, as for Fig.~\ref{tempbar}.
}
\label{metbar}  
\end{figure}

\subsection{Abundance averages}
\label{zav}

Figures~\ref{metiso} and \ref{metbar} show the trend of $\ZicmM$ and
$\ZicmL$ as a function of $\Tz/\Tvir$ (for quasi-isothermal and
quasi-polytropic models) or $r_g/\rc$ (for TMB models), for the same
total mass $M$ and potentials used in the previous figures, and for
the abundance profile~(\ref{degra}), with the parameter values
specified below that equation.  These figures show that $\ZicmL$ (red
lines) always overestimates $\ZicmM$ (black lines), a result of the
larger weight that central regions have in the calculation of
$\ZicmL$.  For all models $\ZicmM$ and $\ZicmL$ decrease with
$\Tz/\Tvir$ or $\rg/\rc$ increasing; this is explained by the fixed
metallicity profile coupled with gas density profiles that become
flatter (Figs.~\ref{iso} and~\ref{poly}), so that the central regions
where the abundance is highest become less and less important in the
integrals of eqs.~(1) and (\ref{eqzicml}) (except for the TMB
$\alpha=2$ models, where the density profile is independent of
$\rg/\rc$).  The steeper density profiles are those of the isothermal
models with $\Tz<2 \Tvir$ (Fig.~\ref{iso}), therefore the decrease of
$\ZicmM$ and $\ZicmL$ at increasing $\Tz$ is more pronounced
in this range of temperatures.

In general, the details of the discrepancy between $\ZicmM$ and
$\ZicmL$, and its trend with $\Tz/\Tvir$ or $r_g/\rc$, depend on how
the density, temperature and abundance profiles differ from each
other.  The overestimate obtained by using luminosity weighted
abundances is stronger for steeper gas density profiles.  For example,
the $\gamma =0$ models have a slightly flatter density profile at the
center than the $\gamma =1$ ones, so that the discrepancy between
$\ZicmM$ and $\ZicmL$ is in general slightly smaller for $\gamma =0$
than for $\gamma =1$.  A similar behavior is presented by TMB
models. The profiles in this family may have a drop in temperature at
the center, and consequently a steep increase in density, that is more
pronounced for $\alpha=2$ (Fig.~\ref{baro}).
Figure~\ref{metbar} reflects this fact, showing the largest
overestimate of $\ZicmL$ among TMB models for $\alpha=2$, while the
smallest is that of $\alpha=0$ models.

In summary, considering all our models $\ZicmL/\ZicmM$
lies in a quite small range: $1.1\lsim \ZicmL/\ZicmM\lsim 1.6$ for
quasi-isothermal models, $1.25\lsim \ZicmL/\ZicmM \lsim 1.45$ for
quasi-polytropic models, and $1.3\lsim \ZicmL/\ZicmM \lsim 1.7$ for
TMB models.

\subsubsection{Changing the dark mass amount and shape}

For all models presented in this paper it can be proved that $\ZicmM$
is independent of $M$, while it depends on $\Tz/\Tvir$ or $\rg/\rc$,
and $\epsilon$ and $\eta$. As a consequence, the values of $\ZicmM$ in
Figs.~\ref{metiso}--\ref{metbar} keep the same for all dark masses $M$
when the metallicity distribution (23) is used with the specified
parameters.  $\ZicmL$ remains almost identical, for $M\geq 3\times
10^{14} M_{\odot}$, with the largest variation for the highest
$\Tz/\Tvir$ and $\rg/\rc$ of just 2\%.  These small variations are
accounted for by the fact that the gas emissivity is not a pure
power-law in temperature.

As for the temperatures, we also investigated the effect of flattening
of the mass distribution. $\ZicmM$ and $\ZicmL$ become smaller when
increasing the flattening with respect to the spherical case, which is
explained by more and more gas mass being displaced at larger
distances from the center, where the abundance is lower. When changing
the values of $\epsilon$ and $\eta$ from zero to (0.5, 0.5), for the
range of $\Tz/\Tvir$ of Fig.~\ref{metiso} and of $\rg/\rc$ of
Fig.~\ref{metbar}, $\ZicmM$ decreases by $\lsim 16$\%, and $\ZicmL$ by
$\lsim 14$\%.  Smaller variations are obtained for more reasonable
flattenings, as for example of the order of $\sim 5$\% for $\epsilon$
and $\eta$ equal to 0.1, 0.3.


\section{Summary and conclusions}
\label{secdis}

In this work we have compared the values of mass and luminosity
weighted metallicity and temperature, for a large set of hydrostatic
gas distributions, some of which resemble those typical of the
intracluster and intragroup media.  In addition, we also computed the
temperature that would be derived from observed spectra (the so-called
spectroscopic-like temperature) by using two recently proposed methods
for its estimate.  The results of this analysis are useful for distant
groups/clusters, or in general for systems with a low number of
observed counts, where only global average values can be recovered
from observations.

This study is based on a few steps. First, the potential well of
triaxial dark matter halos, with different density slopes and
adjustable flattenings, was built analytically by means of homeoidal
expansion, which gives a simple yet accurate analytical approximation
of the true potential.  In the second step we showed how to construct
hydrostatic analytical solutions for triaxial truncated density
distributions, and presented the equilibrium configurations in the
quasi-isothermal and quasi-polytropic cases, and for a family of
modified $\beta$ models.  In the third step we superimposed a
metallicity distribution derived from observations of the
ICM/IGM. Finally, the gas radiative properties were computed by using
the cooling function $\Lambda(T,Z)$ appropriate for our range of gas
temperatures and a chosen sensitivity band of 0.3--8 keV.  Mass and
luminosity weighted temperature and abundances for the models were
then obtained, thanks to the Projection Theorem.

The main results can be summarized as follows.

\begin{itemize}

\item The quasi-isothermal and polytropic models show gas density and
  temperature profiles similar to those observed for non-cool core
  clusters, while those of TMB models with $\alpha=1$ or $\alpha=2$
  resemble cool-core clusters.  In particular the temperature profiles
  of the latter TMB models, when rescaled to $T_{500}$, compare well
  in shape and normalization with observed profiles.  In general,
  $\TicmM/T_{500}$ ranges between 1 and 1.5, and it is 1.3 for TMB
  models with $\alpha=2$.

\item The luminosity-weighted temperature $\TicmL$ overestimates
  $\TicmM$ up to a factor of $\sim 2$, and the discrepancies increase
  with increasing gas temperature (scaled by $\Tvir$) for
  quasi-isothermal and polytropic models, or for increasing $\rg /
  \rc$ for TMB models. For these latter models with $\alpha=2$ the
  overestimate is milder (a factor of $\simeq 1.3$).

\item $\Tspec $ always provides a less serious overestimate of
  $\TicmM$ than $\TicmL$.  The discrepancy bewteen $\Tspec$ and
  $\TicmM$ becomes smaller for increasing $\Tz/\Tvir$ and for
  increasing $\rg/\rc$.

\item The exception to a general overestimate of $\TicmM$ is that of
  "cold" ($\Tz\lsim 1.2\Tvir$) quasi-isothermal models, where the
  three temperatures $\TicmM$, $\TicmL$ and $\Tspec$ are very
  similar. Also, $\TicmL$ and $\Tspec$ keep close up to $\Tz\simeq
  2\Tvir$, and depart for higher $\Tz/\Tvir$.

\item When changing the total dark mass $M$, the general behavior of
  $\TicmM$, $\TicmL$ and $\Tspec$ described above remains the
  same. The values of $\TicmL/\Tvir$ and $\Tspec/\Tvir$ turn out to
  keep within $ 5$\% by changing $M$, for the range of masses typical
  of large groups and clusters ($M \geq 3\times
  10^{14}M_{\odot}$). $\TicmM/\Tvir$ is instead independent of $M$.

\item In the explored range of triaxiality, flattening effects are not
  strong: the average temperatures normalized to $\Tvir$ remain the
  same within 5\%.

\item The only exception to the small ($\lsim 5$\%) variance with a
  change of shape or mass is given by the "cool-core" models (TMB
  models with $\alpha=2$): the increase of $\Tspec/\Tvir$ can be as
  large as 13\% going from $M=3\times 10^{14} M_{\odot}$ to
  $M=1.4\times 10^{15} M_{\odot}$, or from spherical to
  $\epsilon=\eta=0.5$ at fixed $M$.

\item The luminosity weighted $\ZicmL$ overestimates the mass weighted
  average abundance $\ZicmM$.  For quasi-polytropic and
  quasi-isothermal models with $\Tz\geq 2.6\Tvir$ we found that
  $1.3\lsim \ZicmL/\ZicmM\lsim 1.5$. This ratio extends over a larger
  range ($1.1\lsim \ZicmL/ \ZicmM\lsim 1.6$) for colder
  quasi-isothermal models ($\Tz\lsim 2\Tvir$).  TMB models show their
  smallest overestimate ($\simeq 1.4$) for $\alpha=0$, and the largest
  ($\simeq 1.7$) for $\alpha=2$, that repoduces the case of "cool
  core" ICM/IGM.

\item Similarly to what found for $\TicmM$, an $M$ variation has no
  effect on $\ZicmM$, and a negligible effect on $\ZicmL$.  The effect
  of flattening is present, but it is not very important.  For
  $(\epsilon,\eta)$ equal to (0.1, 0.3) and (0.5, 0.5), $\ZicmM$ and
  $\ZicmL$ decrease by $\leq $few \%, and by $\leq 13$\% respectively.

\end{itemize}

Thus, we have shown that when deprojection is not feasible or robust
(as in the case of distant objects, significant deviations from
spherical symmetry, etc.), the alternative approach of considering the
global average values of temperature and abundance, obtained as
surface integrals over the image, has the advantage over deprojection
of being independent of the shape of the system and of the relative
orientation to the observer, but in presence of non-uniform
metallicity and temperature distributions it must be calibrated by
computing the appropriate correcting factors, as those determined in
this paper. It would be interesting to apply both methods
(deprojection vs. surface average) to real systems with detailed
observations, and to compare the results.

\section*{Acknowledgments}
We thank S. Ettori and E. Rasia for comments and the referee for a
constructive report. We are grateful to A. Vikhlinin for providing the
tabulated values required for the estimate of the spectroscopic
temperature.

\appendix
\section{Analytical results}
\subsection{The Projection Theorem}

The numerical integrations performed in this paper are based on a very
simple but far-reaching mathematical identity holding between volume
and surface integrals of projected properties of ``transparent''
systems of general shape (e.g., see Ciotti 2000).  In the present
context, let us consider a property $P(\xv)$ (for example the ICM
temperature or metallicity), associated with a field $\nu(\xv)$ (for
example the ICM emissivity) acting as a ``weight''.  Without loss of
generality, we can suppose the line-of-sight to coincide with the
$z$-axis, so $(x,y)$ is the projection plane.  The weighted
projection is naturally defined as
\[
\Sigma(x,y) \times P_{\rm pr}(x,y)\equiv \int_{-\infty}^{\infty}\nu (\xv)
\,P(\xv)\,dz,
\]
where $\Sigma(x,y)=\int_{-\infty}^{\infty}\nu (\xv)\,dz$.  
From the identity $\int\Sigma\times P_{\rm pr}\,dx\,dy=\int\nu
(\xv)\,P(\xv)\,d^3\xv$, it follows that
\[
<P>_{\nu}\equiv\frac{\displaystyle\int\Sigma\times P_{\rm pr}\,dx\,dy}
                    {\displaystyle\int\Sigma \,dx\,dy}
              =\frac{\displaystyle\int\nu(\xv)\,P(\xv)d^3\xv}
                    {\displaystyle\int\nu(\xv)\,d^3\xv},
\]
where the integrals that appear in this definition extend on the whole
image.  In practice, the surface weighted integral of a projected
property coincides with the volume weigthed integral of this property;
in turns, this proves that $<P>_{\nu}$ is independent of the viewing
angle.

\subsection{The density approach}

As well known, for a gas in hydrostatic equilibrium in a potential
well $\phi(\xv)$
\[
\nabla p=-\rho\nabla \phi ,
\]
and the density, temperature and pressure distributions are stratified
on the equipotential surfaces, i.e., $\rho=\rho(\phi)$, $p=p(\phi)$,
and $T=T(\phi)$, so that the gas is barotropic.  
In standard applications some relation $p(\rho)$ is assigned,
and eq.~(A3) is solved. In the density approach instead the
functional form $\rho(\phi)$ is assigned. In this case it is
trivial to prove that the function
\[
H(\phi)\equiv\int_{\phi}^0 \rho(t)\,dt
\]
satisfies the identity $\nabla H=-\rho(\phi)\nabla\phi$, and so
integrating eq.~(A3) over an arbitrary path it follows that
\[
p(\xv)-p_*=H[\phi(\xv)]-H[\phi_*],
\]
where $p_*$ and $\phi_*$ are the pressure and potential values at some
reference position $\xv_*$. For untruncated models in halos of finite
mass the natural choice is to set $p_*=0$ and $\phi_*=0$ at
$\xv_*=\infty$. For models as those considered in Sect. 2.2, with a
finite value of the central potential $\phiz$ and a truncated density
$\rho-\rhot$, where $\rhot=\rho(\phit)$, it is natural to set
$p_*=p_t=0$, so that eq.~(A5) becomes
\[
p(\xv)=|\phiz |
       \left[\tilde H(\Psi)-\tilde H(\Psit) - 
             \rhot\times (\Psi -\Psit)\right ],
\]
where $\Psi\equiv\phi/\phiz$ and $\tilde H(\Psi)
=\int_0^{\Psi}\rho(\Psi)d\Psi$.

The simple idea behind the density approach is to choose a spherical
gas density distribution $\rho(r)$ with the desired radial profile,
and a spherical potential $\phi(r)$. Due to the monotonicity of the
potential (guaranteed by Gauss theorem), it is always possible (in
principle) to eliminate the radius between the two distributions, thus
obtaining $\rho(\phi)$, and the function $H$.  In the second step one
deform the spherical potential (for example by using homeoidal
expansion as in this paper, or with the complex shift method described
in Ciotti \& Giampieri 2007, or by functional substitution as in the
Miyamoto-Nagai 1980 disk case) but assumes that the function
$\rho(\phi)$ is the same as in the spherical case. As a consequence,
the function $H$ is still exact, while the deformed density profile is
more and more similar to the spherical initial distribution for
smaller and smaller potential deformations.

In the family of models presented in Sect. 2.2.3, remarkably 
simple explicit cases are obtained for
$\beta=2/3$ and $\alpha=0,1,2$.  In particular, after defining $\Psi$
as in equation (15), for $\alpha =0$ we obtain
\begin{eqnarray}
{\tilde H(\Psi)\over\rhoz }&=&{b^2\,\Psi\over b^2+1}+
                            {b(1-b^2)\over (b^2+1)^2}
                               {\rm arctan}{b\Psi\over 1-\Psi}+\cr
                            && {b^2\log[(1-\Psi)^2+b^2\Psi^2]\over
                                (b^2+1)^2} ,
\end{eqnarray}
for $\alpha=1$
\begin{eqnarray}
{\tilde H(\Psi)\over\rhoz }&=&b^2{1-\sqrt{(1-\Psi)^2+b^2\Psi^2}\over
                                   b^2+1}+\cr
                            && b\log{b\psi+\sqrt{(1-\Psi)^2+b^2\Psi^2}\over
                                     1-\Psi}+
                               {b^2(2+b^2)\over (b^2+1)^{3/2}}\times\cr
                            && 
\log{{1-(1+b^2)\Psi +\sqrt{1+b^2}\sqrt{(1-\Psi)^2+b^2\Psi^2}}\over
                                    1+\sqrt{1+b^2}} ,
\end{eqnarray}
and finally 
\[
{\tilde H(\Psi)\over\rhoz}=b^2{\Psi (2-\Psi)\over 1-\Psi}+
2b^2\log (1-\Psi)
\]
for $\alpha=2$. 

It can be easily proved, for example by series expansion 
and by order matching of the hydrostatic equation near the origin that 
the equilibrium temperature for $r\to 0$ converges to a positive
value for $\alpha=0$, 
while for $\alpha >0$ it vanishes as $r^{\alpha}$ ($0<\alpha<1$), as
$-r\log r$ ($\alpha =1$), and finally as $r$ ($\alpha >1$).  In
particular, in the $\alpha=2$ models, the vanishing of temperature
near the center is not strong enough to compensate the density
(square) increase as $r^{-4}$, and this leads to a divergent
denominator in the definition of $\Tspec(\delta)$ for $\delta <5/2$.

\section{The cooling function}

The functions $\Lambda (T,0)$ (in units of $10^{-24}$ erg cm$^3$
s$^{-1}$) and $g(T)$ (dimensionless) appearing in eq.~(\ref{eqlambda})
describe the cooling function over the 0.3--8 keV band for plasmas in
the temperature range $kT=0.1-16$ keV and with the abundance ratios of
Grevesse \& Sauval (1998).  In order to speed up the numerical
integrations, in the code we use the following non-linear fits 
(the Pad\'e approximants) of the
numerical values produced with APEC
\[
\Lambda(T,0) = {a+ b T + c T^2+d T^3 +e T^4\over
                f+ g T + h T^2+i T^3 + l T^4 + m T^5} ,
\]
and
\[
g(T) = {a+ b T + c T^2+d T^3 +e T^4 + f T^5 + g T^6\over
        h+ i T + l T^2+m T^3 + n T^4 + o T^5 +p T^6}.
\]
In the expressions above $T$ is expressed in keV, and the
coefficients are given in Table B1.

\begin{table}
\centering
\begin{minipage}{140mm}
\caption{Double-precision coefficients for the numerical cooling function}
\begin{tabular}{@{} l c c @{}}
\hline
\noalign{\smallskip}
\hline
\noalign{\smallskip}
  & $\Lambda(T,0)$ & $g(T)$ \\
\noalign{\smallskip}
\hline
\noalign{\smallskip}
$a$ & 0.006182527125052084 & 70.36553428793282 \\
$b$ & -0.19660367102929072 & -445.6996196648889 \\
$c$ & 2.0971400804256      & 1154.002954067754 \\
$d$ & -8.007730363728133   & -334.98255766589193 \\
$e$ & 11.575056847530487   & 232.9431532001086 \\
$f$ & -0.005867677028323465& 185.87403547238424 \\
$g$ & 0.1915050008268439   & -6.733225934810589 \\
$h$ & -0.8925516119292456  & -0.5099502967624234 \\
$i$ & 1.2939305724619414   & 39.52256435778057 \\
$l$ & 0.4960145169989057   & -357.9367436666503 \\
$m$ & 0.010460898901591362 & 1343.6559752815178 \\
$n$ & & -1877.1810281439414 \\
$o$ & & 1208.8015587434686 \\
$p$ & & -27.997494448025837 \\
\noalign{\smallskip}
\hline
\end{tabular} 
\end{minipage}
\end{table}
\end{document}